\documentclass[12pt]{article}
\usepackage[margin=1in]{geometry}
\usepackage{amsfonts,amssymb,epsfig,amsmath}
\usepackage{slashed}
\usepackage{color}
\usepackage{hyperref}

\renewcommand{\text}[1]{#1}

\newcommand{\be}{\begin{equation}}
\newcommand{\ee}{\end{equation}}
\newcommand{\ben}{\begin{displaymath}}
\newcommand{\een}{\end{displaymath}}
\newcommand{\bea}{\begin{eqnarray}}
\newcommand{\eea}{\end{eqnarray}}
\newcommand{\bean}{\begin{eqnarray*}}
\newcommand{\eean}{\end{eqnarray*}}
\newcommand{\nn}{\nonumber \\}
\newcommand{\ba}{\begin{array}}
\newcommand{\ea}{\end{array}}
\newcommand{\bi}{\begin{itemize}}
\newcommand{\ei}{\end{itemize}}

\def\G{\Gamma}

\def\G{\Gamma}

\def\e{\epsilon}
\def\s{\sigma}
\def\e{\epsilon}

\DeclareMathOperator{\vol}{vol}

\newcommand{\dd}{\mathrm{d}}
\newcommand{\DD}{\mathrm{D}}

\begin{document} 

\begin{titlepage}

\vfill
\begin{flushright}
FPAUO-14/07 \\
DMUS-MP-14/12 \\
YITP-SB-1432
\end{flushright}

\vfill

\begin{center}
   \baselineskip=16pt
   {\Large \bf Supersymmetry and non-Abelian T-duality \\  in type II supergravity}
   \vskip 2cm
     \"Ozg\"ur Kelekci$^a$, Yolanda Lozano$^b$,  Niall T. Macpherson$^c$, Eoin \'O Colg\'ain$^{d, e}$
       \vskip .6cm
             \begin{small}
               \textit{$^a$ Physics Department \& Faculty of Engineering and Architecture, Siirt University, \\ Kezer/Siirt 56100, TURKEY}
                 \vspace{3mm}
                 
                 \textit{$^b$Departamento de F\'isica, 
		 Universidad de Oviedo, 
33007 Oviedo, SPAIN}
                 \vspace{3mm} 
                 
                 \textit{$^c$ Department of Physics, Swansea University Singleton Park, Swansea SA2 8PP, UK}
                 \vspace{3mm} 
                 
                 \textit{$^d$ C.N.Yang Institute for Theoretical Physics, SUNY Stony Brook, NY 11794-3840, USA}
                 \vspace{3mm}
                 
                  \textit{$^e$ Department of Mathematics, University of Surrey, Guildford GU2 7XH, UK}
                 \vspace{3mm}

             \end{small}
\end{center}

\vfill \begin{center} \textbf{Abstract}\end{center} \begin{quote}
We study the effect of T-duality on supersymmetry in the context of type II supergravity. For both U(1) Abelian and SU(2) non-Abelian T-duality, we demonstrate that the supersymmetry variations after T-duality are related to the variations before T-duality through the Kosmann spinorial Lie derivative, which vanishes when the Killing spinors are independent of the T-duality directions. As a byproduct of our analysis, we present closed expressions for SU(2) T-duality in a class of spacetimes with diagonal Bianchi IX symmetry and comment on specific examples of T-dual geometries, including a novel AdS$_3$ geometry with large $\mathcal{N} = (0,4)$ superconformal symmetry. 
 \end{quote} \vfill

\end{titlepage}


\tableofcontents
\section{Introduction}

(Abelian) T-duality has been very successfully used over the years as a solution generating technique in string theory. Being a genuine symmetry of string theory, both in its perturbative and low energy expansions, the newly generated solution is guaranteed to produce a consistent string theory background, which should exhibit the same basic properties as the original one. Still, puzzling situations arise when the dual theory is explored in its low energy limit, such as the emergence of singularities and the explicit breaking of spacetime supersymmetries \cite{Bakas:1994ba, Bakas:1995hc, Bergshoeff:1994cb, Duff:1998us}. It is well-known that those symmetries of the original theory, non-commuting with the ones that generate the duality, and supersymmetry is no exception to it, are only realized non-locally in the dual \cite{Kiritsis:1993ju}, and are thus typically non-manifest in the low energy expansion.

The non-Abelian cousin of Abelian T-duality was first explored in the 90's at the level of the string sigma model \cite{de la Ossa:1992vc}. Although very similar at this level to its Abelian counterpart, this transformation did not reach the status of a string theory symmetry for two main reasons. First, the derivation in \cite{de la Ossa:1992vc} only works at tree level in string perturbation theory, and second, it is only known to respect conformal symmetry to first order in $\alpha^\prime$. Yet, starting with the derivation of the transformation rules of the RR potentials by Sfetsos and Thompson in \cite{Sfetsos:2010uq}, this transformation has been very successfully used in recent years to generate new supergravity backgrounds with putative CFT duals, which in many cases share striking similarities with CFTs derived in the literature in completely different set-ups. Interesting examples are the relation found in \cite{Sfetsos:2010uq} between an SU(2) dual of AdS$_5\times$ S$^5$  and the $\mathcal{N}=2$ Gaiotto-Maldacena geometries \cite{Gaiotto:2009gz},  its $\mathcal{N}=1$ counterpart, worked out in \cite{Itsios:2012zv, Itsios:2013wd}, which relates an SU(2) dual of Klebanov-Witten \cite{Klebanov:1998hh} with the $\mathcal{N}=1$ CFTs in \cite{Bah:2011vv, Bah:2012dg}, the 
connection between an SU(2) dual  \cite{Lozano:2012au, Lozano:2013oma} of the AdS$_6\times$ S$^4$ Brandhuber and Oz background  \cite{Brandhuber:1999np} and the 5d fixed-point theories classified in \cite{Intriligator:1997pq}, or the explicit realization of the $\mathcal{N}=2$ 3d CFT that must be dual to the new AdS$_4$ background constructed in \cite{Lozano:2014ata}. 
These examples show that, in contrast with its Abelian counterpart, non-Abelian T-duality may produce AdS/CFT pairs with very different properties from the original ones.

What is clear is that non-Abelian T-duality is a powerful solution generating technique on par with Ehlers or Geroch transformations in pure gravity. Although it has been explored largely on a case-by-case basis, for example \cite{Sfetsos:2010uq,  Itsios:2013wd, Lozano:2012au, Lozano:2014ata, Sfetsos:2014tza}, it has been proved that the simplest form of non-Abelian T-duality, namely a left or right-acting SU(2) transformation, is a symmetry of the equations of motion (EOMs) of type II supergravity for spacetimes with SO(4) isometry \cite{Itsios:2012dc}. In this paper, we extend the analysis to a much larger class of spacetimes with Bianchi IX symmetry and, eschewing the EOMs - partial results for a class of spacetimes with SU(2) isometry can be found in \cite{Jeong:2013jfc} -  show how non-Abelian T-duality transforms the supersymmetry conditions of type II supergravity  \footnote{This covers specific examples where supersymmetry has been studied in the context of $G$-structures \cite{Barranco:2013fza, Macpherson:2013zba, Gaillard:2013vsa, Caceres:2014uoa}. }.  Indeed, since supersymmetry implies the Einstein and dilaton equations under mild assumptions, by combining our results with existing integrability results \cite{Lust:2004ig, Gauntlett:2005ww} \footnote{Results can be extended to calibrated D-branes in a non-supersymmetric setting \cite{Lust:2008zd}.}, it should be feasible to show that non-Abelian T-duality is a generic solution generating technique in type II supergravity. We leave this direction to future work. 

A further motivation for this work is to derive, in an alternative way, the transformation for the Ramond-Ramond (RR) sector under a non-Abelian T-duality transformation. We recall that the transformation for the RR sector under non-Abelian T-duality, which was originally proposed as a transformation on the flux bispinor using the relative twist of left and right movers \cite{Sfetsos:2010uq}, can be derived via dimensional reduction \cite{Itsios:2012dc, Jeong:2013jfc} and can be inferred from the Fourier-Mukai transformation \cite{Gevorgyan:2013xka}. Here we offer an alternative derivation using the gravitino supersymmetry variation, which largely parallels the Abelian calculation \cite{Hassan:1999bv}. It would be interesting to recover similar results by studying vertex operators \cite{Polchinski:1996na} or pure spinor formalism \cite{Benichou:2008it}, approaches which have led to the desired result in the Abelian case.

The general treatment of supersymmetry in this work also raises the possibility that non-Abelian T-duality may be useful for identifying new supersymmetry-preserving integrable-deformations in the context of AdS/CFT.  It is known that the application of a sequence of (Abelian) T-duality - shift - T-duality (TsT) transformations leads to integrable deformations of AdS$_5 \times$ S$^5$ \cite{Lunin:2005jy, Frolov:2005dj}. Recently, it has been shown that non-Abelian T-duals arise as limits of integrable theories \cite{Sfetsos:2013wia} (see also \cite{Hollowood:2014rla, Itsios:2014vfa}), further implying that non-Abelian T-duality preserves integrability. Some hint of this can be found in studies of semi-classical strings in T-dual geometries \cite{Zacarias:2014wta, Pradhan:2014zqa}, and it is expected that the methods of \cite{Basu:2011di}, recently applied to the related complex $\beta$-deformation of $\mathcal{N}=4$ super-Yang-Mills  \cite{Giataganas:2013dha}, will not observe chaotic behaviour \footnote{We thank D. Giataganas for discussion on this point.}. It would be interesting to see if analogous TsT transformations can be found for non-Abelian T-duality.

In this paper we take the first steps towards exploring the issue of spacetime supersymmetry versus non-Abelian T-duality in a general setting. Given the possibility of non-locally realized supersymmetries at the worldsheet level \footnote{Worldsheet supersymmetry for both Abelian and non-Abelian T-duality has been discussed in \cite{Bakas:1995hc, Alvarez:1995np, Sfetsos:1996pm}.}, we will resort to exploring the leading term in the low energy expansion. This issue was explored at this level in the Abelian set-up in \cite{Bergshoeff:1994cb}, where it was shown for ten-dimensional $\mathcal{N}=1$ supergravity that the Killing spinors had to be independent of the T-duality direction for supersymmetry to be unaffected by the duality. In this paper we will extend this Abelian T-duality result to type II ($\mathcal{N}=2$) supergravity, and generalise it further to SU(2) non-Abelian T-duality in a (diagonal) Bianchi IX family of spacetimes. We will uncover a unifying picture. For both Abelian and non-Abelian T-duality, we will demonstrate that the supersymmetry conditions for the T-dual geometry can be written in terms of the supersymmetry conditions of the original geometry up to additional Kosmann spinorial Lie derivative \cite{Kosmann} terms that vanish when supersymmetry is preserved. Therefore the Kosmann derivative encapsulates all information about supersymmetry breaking at leading order in the low energy expansion and provides an alternative way to determine preserved supersymmetry without an explicit knowledge of the Killing spinors of the original geometry. 

The paper is organized as follows. Section 2 summarizes the Abelian case, emphasizing the role played by the Kosmann derivative. As the path is well-trodden, we restrict our analysis to massive IIA. In Section 3 we start our generalization to the SU(2) case \footnote{In non-Abelian T-duality, the chirality of the T-dual theory depends on the dimension of the isometry group. SU(2) having three generators, flips chirality, just as in the U(1) case.}. In this section we explore the Kosmann derivative in spacetimes with SU(2) isometry, to later, in section 4, introduce the NS sector of the SU(2) dual background and identify a preferred T-dual frame. In section 5 we work out the supersymmetry variations, and show that they are the same before and after duality  provided the Killing spinors do not depend on the SU(2) directions in our preferred frame. We show that this condition is equivalent to the vanishing of the Kosmann derivative. We also generate the dual RR fluxes by imposing the matching of the gravitino supersymmetry variations and show that it is consistent with the remaining supersymmetry variations and the expected Bianchi identities. In section 6 we briefly discuss the extensions of the previous results to type IIB in non-Abelian cases. Section 7 contains some examples that illustrate the main results in the paper. These concern the SU(2) T-dual of the AdS$_5\times$ Y$^{p,q}$ geometries recently constructed in \cite{Sfetsos:2014tza}, the SU(2) T-dual of the AdS$_4\times \mathbb{CP}^3$ background \cite{Lozano:2014ata}, and the new SU(2) T-dual of an AdS$_3\times$ S$^3\times$ S$^3\times \mathbb{R}$ geometry \cite{de Boer:1999rh} that we work out in this section. Section 8 contains our Conclusions, where we summarize the main results of the paper and discuss further prospects. Appendix A contains a brief account on squashed S$^3$ (Bianchi IX) spacetimes and Appendix B some technical details needed for the computation of the dual supersymmetry variations.

\section{Abelian case}
\label{sec:abelian}
\setcounter{equation}{0}

In this section we illustrate what happens to spacetime supersymmetry under Abelian T-duality. It is widely reported in the literature that if the Kosmann spinorial Lie derivative \cite{Kosmann} vanishes, then supersymmetry will be unaffected by T-duality. A statement along these lines exists for type I supergravity in ten dimensions \cite{Bergshoeff:1994cb} and it is expected to generalise to type II. Here we will show that if the Kosmann derivative vanishes, then supersymmetry is preserved under T-duality. As an immediate corollary, it follows that the Kosmann derivative determines the extent to which supersymmetry is broken when it is non-zero. 

For concreteness, we confine our attention to a T-duality from massive IIA, but expect there to be no difference in the analysis in the context of type IIB. Much of the analysis here parallels that of \cite{Hassan:1999bv}, with the notable difference that we will be interested in the role of the Kosmann derivative, which will involve making it explicit in our supersymmetry analysis. 

We consider a general spacetime with a U(1) isometry: 
\bea
\label{original_spacetime}
\dd s^2_{10} &=& \dd s^2_9 + e^{2 C} (\dd z +A_1)^2, \nn 
B_2 &=& B + B_1 \wedge \dd z, \nn
F_{0} &=& m, \nn
F_{2} &=& G_2 + G_1 \wedge (\dd z + A_1), \nn
F_{4} &=& G_4 + G_3 \wedge (\dd z + A_1), 
\eea
where $m$ is the Romans' mass and we have allowed for various fields $C, A_1, B_1$ and $G_p$, $p=1, 2, 3, 4$, which are defined on the transverse metric $\dd s^2_9$. In addition, we consider a non-zero scalar dilaton, $\Phi$ and $B$ is the transverse NS two-form, which will play no role in the transformation. A short calculation produces the spin connection for this spacetime: 
\bea
\label{abelian_spin}
\omega^{\mu}_{~\nu} &=& \bar{\omega}^{\mu}_{~\nu} - \frac{1}{2} e^{C} F^{\mu}_{~ \nu} e^{z}, \nn
\omega^{z}_{~\mu} &=& \partial_{\mu} C e^{z} + \frac{1}{2} e^{C} F_{\mu \rho} e^{\rho}, 
\eea
where $F= \dd A_1$ and $\mu = 0,...,8$ denotes transverse directions. We are working in the natural frame where 
\begin{equation}\label{eq: TDframe}
e^{\mu} = \bar{e}^{\mu},~~~ e^{z} = e^{C} (\dd z +A_1) . 
\end{equation}

The Lie derivative of a spinor with respect to a Killing direction $K$, or alternatively Kosmann derivative,  may be defined as \cite{Kosmann} 
\be
\label{Kosmann}
\mathcal{L}_{K} \eta = K^{a} \nabla_{a} \eta + \frac{1}{8} (\dd K)_{a b} \Gamma^{a b} \eta, 
\ee
where $\nabla_{a} \equiv \partial_{a} + \frac{1}{4} \omega_{abc} \G^{bc}$. Adopting $K = \partial_{z}$, it is an easy exercise to show that the spin connection and $\dd K$ component cancel, leading to the relationship 
\be
\label{identity}
\mathcal{L}_{\partial_{z}} \eta = \partial_{z} \eta. 
\ee
As a result of this identity the Kosmann derivative with respect to $\partial_{z}$ vanishes when the Killing spinors do not depend on the coordinate $z$
\be
\mathcal{L}_{\partial_{z}} \eta = 0 \Leftrightarrow \partial_{z} \eta = 0. 
\ee

To extract a projection condition from the Kosmann derivative, use may be made of the gravitino variation - see later (\ref{gravitino_IIA}) - in the $z$-direction to substitute for the covariant derivative \footnote{Where convenient we use slashed notation, where $\slashed{A} = A_{\mu_1 ...\mu_p} \G^{\mu_1 ...\mu_p}$ for a p-form $A$.} : 
\bea
\label{Kosmann_rewrite}
\mathcal{L}_{\partial_z} \eta &=& e^{2 C} \frac{1}{8} \slashed{F} \eta + \frac{1}{2} e^{C} \slashed{\partial} C \G^{z} \eta + \frac{1}{8} (\dd B_1)_{\mu \nu} \G^{\mu \nu} \sigma^3 \eta - \frac{1}{8} e^{C+ \Phi} \biggl[  m \sigma^1\nn  &+& \frac{1}{2} \slashed{G}_2 i \sigma^2 + e^{-C} \slashed{G}_1 \G^{z} i \sigma^2 + \frac{1}{24} \slashed{G}_4  \sigma^1 + \frac{1}{6} e^{-C} \slashed{G}_3 \G^{z} \sigma^1   \biggr] \G_{z} \eta 
\eea
Here, we have used 
\bea
H_3 &=& \dd B - \dd B_1 \wedge A_1 + \dd B_1 \wedge (\dd z + A_1), \nn
 &=&  H + e^{-C} \dd B_1 \wedge e^{z}. 
\eea

With the help of the T-duality conventions of Hassan \cite{Hassan:1999bv}, one can determine the T-dual geometry: 
\bea
\dd \hat{s}^2_{10} &=& \dd s^2_9 + e^{-2 C} (\dd z +B_1)^2, \nn
\hat{B}_2 &=& B + A_1 \wedge ( \dd z + B_1) \nn
e^{\hat{\Phi}} &=& e^{ \Phi-C}, \nn
\hat{F}_{1} &=&   m (\dd z +B_1) - G_1 , \nn
\hat{F}_{3} &=&  G_2 \wedge (\dd z + B_1) - G_3, \nn
\hat{F}_{5} &=& (1+ *_{10}) G_4 \wedge (\dd z + B_1). 
\eea
To ensure consistency, it is easy to confirm that the Bianchi identities give the same result before and after T-duality. This then is a check on the sign of the RR fluxes. We further notice that $A_1$ and $B_1$ are interchanged and the sign of $C$ is flipped, making it easy to work out the T-dual spin connection from (\ref{abelian_spin}). 

We are now in a position to identify how the supersymmetry variations transform.  We start with some conventions. We recall the supersymmetry variations for type IIA \cite{Itsios:2012dc, Hassan:1999bv}
\bea
\label{dilatino_IIA} \delta \lambda &=& \frac{1}{2} \slashed{\partial} \Phi {\eta} - \frac{1}{24} \slashed{H}_3 \sigma^3 {\eta} + \frac{1}{8} e^{\Phi} \left[5 m \sigma^1 + \frac{3}{2} \slashed{F_2} (i \sigma^2) + \frac{1}{24} \slashed{F}_4 \sigma^1 \right] \eta, \\
\label{gravitino_IIA} \delta \Psi_{M} &=& \nabla_{M} {\eta} - \frac{1}{8} H_{3\,M N P} \Gamma^{N P} \sigma^3 {\eta} + \frac{1}{8} e^{\Phi} \left[ F_0 \sigma^1 + \frac{1}{2} \slashed{F}_2 i \sigma^2 + \frac{1}{24} \slashed{F}_4 \sigma^1 \right] \Gamma_{M} \eta, 
\eea 
and type IIB 
\bea
\label{dilatino_IIB} \delta \hat{\lambda} &=& \frac{1}{2} \slashed{\partial} \Phi \hat{\eta} - \frac{1}{24} \slashed{H}_3 \sigma^3 \hat{\eta} + \frac{1}{2} e^{\Phi} \left[ \slashed{F_1} (i \sigma^2) + \frac{1}{12} \slashed{F}_3 \sigma^1 \right] \hat{\eta}, \\
\label{gravitino_IIB} \delta \hat{\Psi}_{M} &=& \nabla_{M} \hat{\eta} - \frac{1}{8} H_{3 \, M N P} \Gamma^{N P} \sigma^3 \hat{\eta} - \frac{1}{8} e^{\Phi} \left[ \slashed{F}_1 i \sigma^2 + \frac{1}{6} \slashed{F}_3 \sigma^1 + \frac{1}{240} \slashed{F}_5 i \sigma^2 \right] \Gamma_{M} \hat{\eta},  
\eea
where $\eta$ is a Majorana-Weyl spinor
\be
\eta = \left( \begin{array}{c} \e_+ \\ \e_- \end{array} \right).  
\ee
We have introduced hats on the transformed Killing spinors to distinguish them from the Killing spinors before T-duality.

Substituting in our spacetime Ansatz (\ref{original_spacetime}), the transverse gravitino variation before T-duality may be expressed as 
\bea
\delta \Psi_{\mu \pm} &=& \nabla_{\mu} \epsilon_{\pm} + \frac{1}{4} e^{C} F_{\mu \nu} \G^{\nu z} \e_{\pm}  \mp \frac{1}{8} H_{\mu \nu \rho} \G^{\nu \rho}  \e_{\pm}  \mp \frac{1}{4} (\dd B_1)_{\mu \rho} e^{-C} \G^{\rho z} \e_{\pm} \nn 
&+& \frac{1}{8} e^{\Phi} \biggl[ m \pm \slashed{G}_1 e^{-C} \G^{z}  \pm \frac{1}{2} \slashed{G}_2 + \frac{1}{6} \slashed{G}_3 e^{-C} \G^{z}  + \frac{1}{24} \slashed{G}_4  \biggr] \G_{\mu} \epsilon_{\mp}.  
\eea

After T-duality, one can again determine the T-dual gravitino variation, $\delta \hat{\Psi}_{\mu \pm}$,  and one notes that they are not independent, provided we also redefine the Majorana-Weyl spinors, 
\be
\label{spinor_relation}
\delta \hat{\Psi}_{\mu +} = - \G^{z} \delta \Psi_{\mu +}, \quad \delta \hat{\Psi}_{\mu -} = \delta \Psi_{\mu -}, \quad \hat{\epsilon}_{+} = - \G^{z} \epsilon_{+}, \quad \hat{\e}_- = \e_-,   
\ee
The choice of sign in the expressions above can be traced to an overall sign in the RR sector, which can be flipped by changing the sign of the fluxes. Note also that the chirality of $\e_+$ has been changed in the T-duality, as is expected. 

There appears to be one small subtlety in this identification. Although the gravitino variations are structurally the same, the partial derivatives in the covariant derivative may not match. To see this, we write 
\be
\nabla_{\mu} = \partial_{\mu} + \frac{1}{4} \bar{\omega}_{\mu \nu} \G^{\mu \nu} - A_{1\,\mu} \partial_{z}  , \quad \hat{\nabla}_{\mu} = \partial_{\mu} + \frac{1}{4} \bar{\omega}_{\mu \nu} \G^{\mu \nu} - B_{1\,\mu} \partial_{z},  
\ee
which is not guaranteed to be the same when $\partial_{z} \eta \neq 0$. So, even at this early stage we notice that the variations only match when either the Killing spinors are independent of the T-duality direction, or if they are not, when both $A_1$ and $B_1$ are zero. 

We next consider the direction on which we T-dualise. Before T-duality, the gravitino variation in this direction is zero. We have assumed this to rewrite the Kosmann derivative in (\ref{Kosmann_rewrite}). For this reason, the gravitino variation before T-duality may be expressed as 
\be
\delta \Psi_{z} = e^{-C} \left( \partial_z - \mathcal{L}_{\partial_z} \right)  \eta = 0, 
\ee
where we have made use of (\ref{identity}). After T-duality, the variation may be expressed in terms of the Kosmann derivative as  
\bea
\delta \hat{\Psi}_{z +} &=& - e^{C} \G^{z} \partial_{z} \epsilon_{+} + \G^{z} e^{-C} \mathcal{L}_{\partial_z} \e_+, \nn
\delta \hat{\Psi}_{z -} &=& e^{C} \partial_{z} \epsilon_{-} + e^{-C} \mathcal{L}_{\partial_{z}} \epsilon_{-}, 
\eea
where we have used the earlier redefinitions of the spinors (\ref{spinor_relation}). Again, we see that supersymmetry will be preserved provided $\partial_{z} \eta = \mathcal{L}_{\partial_{z}} \eta = 0$.  

Finally, we come to the dilatino variation (\ref{dilatino_IIA}). The dilatino variation before T-duality is 
\bea
\delta \lambda_{\pm} &=& \frac{1}{2} \slashed{\partial} \Phi \e_{\pm} \mp \frac{1}{24} \slashed{H} \e_{\pm} \mp \frac{1}{8} (\dd B_1)_{\mu \nu}  e^{-C} \G^{\mu \nu} \G^{ z} \e_{\pm} \nn
&+& \frac{1}{8} \biggl[  5 m \pm \frac{3}{2} \slashed{G}_2 \pm 3 \slashed{G}_1 e^{-C} \G^{z} + \frac{1}{6} \slashed{G}_3 e^{-C} \G^{z} + \frac{1}{24} \slashed{G}_4 \biggr] \e_{\mp}. 
\eea
Working out the analogous expressions for the T-dual dilatino variation, $\delta \hat{\lambda}$, we identify the following mappings: 
\bea
\delta \hat{\lambda}_+ &=&  \G^{z} \delta \lambda_+  + e^{-C} \mathcal{L}_{\partial_{z}} \epsilon_{+}, \nn
\delta \hat{\lambda}_- &=&  \delta \lambda_-  + e^{-C} \G^{z} \mathcal{L}_{\partial_{z}} \epsilon_{-}, 
\eea
and arrive at the same conclusion as before. 

In summary, starting with the most general Ansatz for a spacetime with U(1) isometry (\ref{original_spacetime}), we have shown that the supersymmetry variations before and after T-duality are the same provided the Killing spinors do not depend on the isometry direction. At some level this statement is intuitively obvious; if supersymmetry does not depend on a direction, it should be clear that we can perform a transformation on this direction and leave supersymmetry unaffected. However, by rewriting the supersymmetry variations, we have noted the special role of the Kosmann derivative. This means one can determine the amount of preserved supersymmetry by extracting a projection condition from the Kosmann derivative, or one can work out the Killing spinor of the original geometry explicitly and identify a subset of the supersymmetries that are independent of the T-duality directions. 

\section{Insights from Kosmann derivative} 
\label{sec:Kosmann}
\setcounter{equation}{0}
Before proceeding to the analogous analysis of the supersymmetry variations for SU(2) T-duality, it is prudent to first study the Kosmann derivative in this context. For spacetimes with SU(2) isometry there is an intrinsic choice in whether one uses left-invariant or right-invariant one-forms to parametrise the SU(2). We will opt to make use of left-invariant one-forms and will work with the metric
\bea
\label{metric}
\dd s^2_{10} &=& \dd s^2_7 + \sum_{a=1}^3 e^{2 C_a} (\tau_a + A^a)^2, 
\eea
where $ \dd \tau_a = \frac{1}{2} \e_{abc} \tau_b \wedge \tau_c$. Explicitly, the left-invariant one-forms may be expressed in terms of coordinates ($\theta, \phi, \psi$) in the accustomed fashion 
\be
\tau_1 = - \sin \psi  \dd \theta + \cos \psi \sin \theta \dd \phi, \quad
\tau_2 = \cos \psi \dd \theta + \sin \psi \sin \theta \dd \phi, \quad
\tau_3 = \dd \psi + \cos \theta \dd \phi.  \nonumber
\ee

Here $A^a$ are SU(2)-valued gauge fields and $C_a$ denote scalar warp factors and generically, when they are all different, the spacetime has a right-acting SU(2) isometry (see appendix \ref{sec:squash}), when two are equal, the symmetry is enhanced to SU(2) $\times$ U(1) and when all three are equal, we recover the SO(4) isometry of a round three-sphere. Gaugings of a round three-sphere in the context of non-Abelian T-duality were considered in \cite{Jeong:2013jfc}. Here we will focus on a more general case. 

Employing a natural orthonormal frame,
\begin{equation}\label{eq: NATDframe}
 e^{\mu} = \bar{e}^{\mu},~~~~ e^{a} = e^{C_a} (\tau_a + A^a),  
\end{equation}
it is a straightforward exercise to work out the spin connection 
\bea
\label{spin}
\omega^{1}_{~2} &=& \frac{1}{2} e^{-C_1 -C_2 -C_3} \left( e^{2 C_1} + e^{2 C_2} - e^{2 C_3} \right) e^3 - \frac{1}{2} e^{-C_1 -C_2} \left( e^{2 C_1} + e^{2 C_2}\right) A^3, \nn
\omega^{1}_{~\mu} &=&  \partial_{\mu} C^1 e^1 -\frac{1}{2} e^{-C_1 -C_2} (e^{2 C_2}-e^{2 C_1}) A^3_{\mu} e^2 + \frac{1}{2} e^{-C_1 -C_3} (e^{2 C_3}-e^{2 C_1}) A^2_{\mu} e^3,  \nn
&+& \frac{1}{2} e^{C_1} F^1_{\mu \rho} e^{\rho}, \nn
\omega^{\mu}_{~\nu} &=& \bar{\omega}^{\mu}_{~\nu} - \sum_{a} \frac{1}{2} e^{C_a} F^{a\mu}_{~~\nu} e^{a}, 
\eea
where we have omitted cyclic terms in $a = 1, 2, 3$. As we have SU(2) gauge fields, it is not surprising to see non-Abelian field strengths, $F^{a} = \dd A^a + \frac{1}{2} \e_{abc} A^b \wedge A^c$, appearing. 

We will now show that the Kosmann derivative with respect to the right-invariant vectors $K_a$ (\ref{right_inv}), dual to the right-invariant one-forms, vanishes when the spinor, $\eta$, is independent of these directions. The right-invariant vectors have the property that they commute with the left-invariant vector fields dual to the left-invariant one-forms parametrising our SU(2) isometry. Equivalently, we could write the metric in terms of right-invariant one-forms and consider the Kosmann derivative with respect to left-invariant vector fields. 

For concreteness, we focus on $K_3$, with similar statements holding for the remaining vectors. The first component in the spinorial derivative may be expressed as 
\be
K_3^{a} \nabla_{a} \eta = \nabla_{\phi} \eta = \left( e^{C_1} \cos \psi \sin \theta \nabla_{1}+ e^{C_2} \sin \psi \sin \theta \nabla_2 + e^{C_3} \cos \theta \nabla_3 \right) \eta, 
\ee
where $\nabla_{\phi}$ denotes derivative with respect to the coordinate $\phi$, while $\nabla_{a}$, $a = 1, 2, 3$, denotes a covariant derivative with respect to left-invariant vectors (\ref{left_inv}).  

Using the metric (\ref{metric}) and the corresponding natural orthonormal frame, we can determine the one-form dual to the vector $K_3$, 
\be
K_3 = \cos \psi \sin \theta e^{C_1} e^{1} + \sin \psi \sin \theta e^{C_2} e^2 + \cos \theta e^{C_3} e^3.  
\ee
Taking derivatives and making use of the following expressions, 
\bea
\dd \theta &=& - \sin \psi (e^{-C_1} e^1-A^1) + \cos \psi (e^{-C_2} e^2-A^2), \nn
\dd \phi &=&  \frac{\cos \psi}{\sin \theta} (e^{-C_1} e^1 - A^1) + \frac{\sin \psi}{\sin \theta} (e^{-C_2} e^2-A^2), \nn
\dd \psi &=& - \cos \psi \frac{\cos \theta}{\sin \theta} (e^{-C_1} e^1 -A^1)-  \sin \psi \frac{\cos \theta}{\sin \theta} (e^{-C_2} e^2 -A^2)\nn  &+& (e^{-C_3} e^3 -A^3),  
\eea
it is possible to show that the $ \dd K_3$ breaks up into three parts 
\bea
\mathcal{L}_{K_3} \eta &=&  \cos \psi \sin \theta e^{C_1} P_1 \eta + \sin \psi \sin \theta e^{C_2} P_2 \eta + \cos \theta e^{C_3} P_3 \eta, \nn
&=& \partial_{\phi} \eta, 
\eea
where 
\bea
P_{1} \eta &=& \nabla_1 \eta  + \frac{1}{8} e^{C_1} \slashed{F}^1  \eta + \frac{1}{4} e^{-C_1 -C_2 -C_3} ( e^{2 C_1} - e^{2 C_2} - e^{2 C_3}) \G^{23}  \eta \nn &+&  \frac{1}{2} \slashed{\partial} C_1 \G^{1}  \eta -\frac{1}{4} e^{-C_1 -C_3} (e^{2 C_1} - e^{2 C_3}) \slashed{A}^2 \G^{3}  \eta
+ \frac{1}{4} (e^{2 C_1} - e^{2 C_2} ) e^{-C_1 - C_2} \slashed{A}^3 \G^{ 2}  \eta, 
\eea
and $P_2 \eta, P_3 \eta$ follow by cyclic symmetry in $1, 2, 3$. Since the spin connection in $\nabla_{a} \eta$ cancels the remaining terms, $P_a \eta$ simply encode left-invariant vectors in our preferred frame acting on the spinor $\eta$.  

One can repeat the above calculation to get similar expressions for $\mathcal{L}_{K_1} \eta$ and $\mathcal{L}_{K_2} \eta$ in terms of $P_{a} \eta$, before inverting to obtain the following: 
\bea
\label{P0}
e^{C_1} P_1 \eta &=& \left( -\sin \phi \sin \psi + \cos \theta \cos \phi \cos \psi \right) \mathcal{L}_{K_1} \eta  - \left( \cos \phi \sin \psi + \cos \theta \sin \phi \cos \psi \right) \mathcal{L}_{K_2} \eta\nn
&+& \sin \theta \cos \psi \mathcal{L}_{K_3} \eta, \nn
e^{C_2} P_2 \eta &=& \left( \sin \phi \cos \psi + \cos \theta \cos \phi \sin \psi \right) \mathcal{L}_{K_1} \eta  + \left( \cos \phi \cos \psi - \cos \theta \sin \phi \sin \psi \right) \mathcal{L}_{K_2} \eta\nn
&+& \sin \theta \sin \psi \mathcal{L}_{K_3} \eta, \nn
e^{C_3} P_3 \eta &=& \sin \theta \left( -\cos \phi \mathcal{L}_{K_1} + \sin \phi \mathcal{L}_{K_2} \right) \eta + \cos \theta \mathcal{L}_{K_3} \eta. 
\eea

It is now easy to convince oneself that 
\be
\mathcal{L}_{K_a} \eta = 0  \Leftrightarrow P_a \eta = 0 \Leftrightarrow \eta~~ \textrm{constant w. r. t.}~~ (\theta, \phi, \psi). 
\ee

As we will see later, the supersymmetry variations after T-duality can be rewritten to make the $P_a \eta$ manifest and they encode supersymmetry breaking during non-Abelian T-duality. More explicitly, focussing on the $K_1$ direction, we get: 
\bea
\label{P1}
P_1 \eta  &=& \frac{1}{8} e^{C_1} \slashed{F}^1 \eta + \frac{1}{4} e^{-C_1 -C_2 -C_3} ( e^{2 C_1} - e^{2 C_2} - e^{2 C_3}) \G^{23} \eta - \frac{1}{4} e^{-C_1 -C_3} (e^{2 C_1} - e^{2 C_3}) \slashed{A}^2 \G^{3} \eta \nn
&+& \frac{1}{4} (e^{2 C_1} - e^{2 C_2} ) e^{-C_1 - C_2} \slashed{A}^3 \G^{ 2} \eta + \frac{1}{2} \slashed{\partial} C_1 \G^{1} \eta  \nn
&+& \frac{1}{8} H_{3 \, 1NP} \G^{NP} \sigma^3 \eta- \frac{1}{8} e^{\Phi} \biggl[ m \sigma^1 + \frac{1}{2} \slashed{F}_2 (i \sigma^2) + \frac{1}{24} \slashed{F}_4 \sigma^1 \biggr] \G_1 \eta,  
\eea
where explicit expressions for $H_3$, $F_2$ and $F_4$ will appear later. Already this is looking considerably more complicated than the Abelian case (\ref{Kosmann_rewrite}), but the expressions for $P_2$, $P_3$ follow from cyclic symmetry in $1, 2, 3$.

\section{Non-Abelian T-duality}
\label{sec:NAT}
\setcounter{equation}{0}

In addition to determining the exact role of the Kosmann derivative in supersymmetry breaking under non-Abelian T-duality, a secondary goal of this work is to provide closed expressions for the SU(2) transformation directly in supergravity. This will provide a needed generalisation of the preliminary results of Ref. \cite{Itsios:2012dc}, which are only valid for spacetimes with SO(4) symmetry. In the future it will be possible to compare the seed solution to our SU(2) Ansatz, read off the data and produce the T-dual geometry. 

In this section, we will focus on the NS sector and leave the analysis of the RR fluxes until the next section, where the transformed RR fluxes will be deduced from the supersymmetry conditions. The SU(2) T-duality rules we use can be found in \cite{Itsios:2013wd}, which we review now. Given a spacetime metric and NS two-form, 
\bea
\dd s^2_{10} &=& G_{\mu \nu} \dd x^{\mu} \dd x^{\nu} + 2 G_{\mu a} \dd x^{\mu} \tau_{a} + g_{ab} \tau_{a} \tau_{b}, \nn
B &=& \frac{1}{2} B_{\mu \nu} \dd x^{\mu} \wedge \dd x^{\nu} + B_{\mu a} \dd x^{\mu} \wedge \tau_a + \frac{1}{2} b_{ab} \tau_a \wedge \tau_b, 
\eea
where $\tau_a$ denote Maurer-Cartan one-forms,  one combines the symmetric and anti-symmetric tensors in the usual way: 
\be
Q_{\mu \nu} = G_{\mu \nu} + B_{\mu \nu}, ~~Q_{\mu a} = G_{\mu a} + B_{\mu a}, ~~ Q_{a \mu} = G_{a \mu} + B_{a \mu}, ~~ q_{ab} =  g_{ab} + b_{ab}.  
\ee
The T-dual symmetric and anti-symmetric tensors then follow from \cite{Itsios:2013wd}, 
\bea
\label{Buscher}
\hat{Q}_{\mu \nu} &=& Q_{\mu \nu} - Q_{\mu a} M_{ab}^{-1} Q_{b \nu}, \quad \hat{q}_{ab} = M_{ab}^{-1}, \nn
\hat{Q}_{\mu a} &=& Q_{\mu b} M_{b a}^{-1}, \quad \hat{Q}_{a \mu} = - M_{ab}^{-1} Q_{b \mu}, 
\eea
where $M_{ab} = q_{ab} + f_{ab}^{~~c} v_c$ is the original tensor in the T-dual directions shifted with respect to the structure constants $f_{abc} $ of the group, further contracted into the Lagrange multiplier $v_c$. We recall that $v_c$ usually impose that the gauge field in the gauging procedure in  \cite{Rocek:1991ps} (see also
\cite{Buscher:1987qj}) is pure gauge, and eventually correspond to the T-dual coordinates.  The dilaton receives a one-loop quantum correction analogous to the Abelian case: 
\be
\hat{\Phi} = \Phi - \frac{1}{2} \ln \textrm{det} M
\ee

Specialising to our SU(2) case, where $\tau_a$ now correspond to left-invariant one-forms and $f_{abc} = \pm \epsilon_{abc}$ \footnote{The sign we use is irrelevant, since we can always change the sign of $v_a$.}, we can identify the symmetric tensors $G_{\mu \nu}$, $G_{\mu a}$ and $g_{ab}$ by comparing to our spacetime Ansatz (\ref{metric}):
\bea
G_{\mu \nu} = g_{\mu \nu} + \sum_{a=1}^3 e^{2 C_a} A^a_{\mu} A^a_{\nu}, \quad  G_{\mu a} = e^{2 C_a} A^{a}_{\mu}, \quad g_{ab} = e^{2 C_a} \delta_{ab}, 
\eea 
where $g_{\mu \nu}$ corresponds to the metric on the transverse seven-dimensional spacetime. In addition, we need to define an NS two-form, which we will take to be 
\be
B_2 = B + (B^a+ \dd b_a) \wedge \tau_a + \frac{1}{2} \epsilon_{abc} b_{a} \tau_{b} \wedge \tau_c, 
\ee
where $B$, $B^a$ and $b_{a}$ are respectively, a two-form, three one-forms and three scalars and we have shifted $B^a$ for convenience. As explained in \cite{Itsios:2013wd}, we can rewrite $B_2$ as 
\be
B_2 = B + B^a \wedge \tau_a + \dd ( b_a \tau_a), 
\ee
making it clear that the contribution from the scalars $b_a$ is a pure gauge contribution \footnote{This Ansatz also precludes a field strength $H_3 \propto \tau_1 \wedge \tau_2 \wedge \tau_3$, a term that is ruled out when one attempts to gauge SU(2) isometries \cite{Plauschinn:2014nha}. }. Thus, $b_a$ will not affect the EOMs and we could set them to zero from the offset. Still, this may preclude solutions, so we retain them in what follows. In the absence of $b_a$, the Lagrange multipliers $v_a$ correspond to T-dual coordinates and get shifted to $z_a = v_a + b_a$ when the B-field is included. For later convenience, we record the field strength in our preferred frame:
\bea
\label{H0}
H_3 &=& H + \DD  B^a \wedge e^{-C_a} e^{a}- \frac{1}{2} \epsilon^{a}_{~bc} e^{-C_b-C_c} B^a \wedge e^{bc}, 
\eea
where we have defined \bea
\DD B^a &=& \dd B^a + \epsilon^{a}_{~bc} A^b \wedge B^c, \nn
H &=&  \dd B - \dd B^a \wedge A^a - \frac{1}{2} \epsilon_{abc} B^a \wedge A^b \wedge A^c. 
\eea
Throughout this work the covariant derivative of an SU(2)-valued p-form, $C^{a}_{p}$ will be uniformly defined to be $\DD C^{a}_{p} = \dd C^{a}_{p} + \epsilon^a_{~bc} A^{b} \wedge C_p^c$.

It is then easy to read off the contribution to the tensors: 
\be
B_{\mu \nu} = B_{\mu \nu}, \quad B_{\mu a} = B^a_{\mu} + \partial_{\mu} b_a, \quad b_{ab} = \epsilon_{abc} b_c. 
\ee

Once the tensors of the original spacetime Ansatz have been determined, the next step is to invert $M_{ab}$
\begin{equation}
M_{ab}^{-1}  =  \frac{1}{\Delta} \left( \begin{array}{ccc}  e^{2(C_2 + C_3)} + z_1^2 & z_1 z_2 - e^{2 C_3} z_3 &   z_1 z_3 + e^{2 C_2 }z_2\\ 
z_1 z_2 + e^{2 C_3} z_3 & e^{2(C_1 + C_3)} + z_2^2 & z_2 z_3 - e^{2 C_1} z_1 \\
z_1 z_3 - e^{2 C_2 } z_2 & z_2 z_3 + e^{2 C_1} z_1 & e^{2 (C_1 + C_2)} +z_3^2
\end{array}\right),  
\end{equation} 
where 
\be
\Delta = e^{2(C_1 + C_2 + C_3)} + e^{2 C_1} z_1^2 + e^{2 C_2} z_2^2 + e^{2 C_3} z_3^2. 
\ee

It is a straightforward task to now read off the T-dual metric from (\ref{Buscher})
\bea
\dd \hat{s}^2 = \frac{1}{\Delta} \biggl[  (z_1 \DD z_1 + z_2 \DD z_2 + z_3 \DD z_3)^2+ e^{2 (C_2 +C_3)} \DD z_1^2 + e^{2(C_1 +C_3)} \DD z_2^2 + e^{2(C_1 +C_2)} \DD z_3^2 \biggr],  \nonumber 
\eea
where the transverse seven-dimensional spacetime comes along for the ride and the covariant derivatives encapsulate the SU(2) gauge fields and the vector components of the original NS two-form
\bea
\DD z_a &=& \dd z_a + B^a - \epsilon_{abc} z_b A^c. 
\eea

The T-dual NS two-form 
\bea
\label{B1}
\hat{B}_2 &=& B- \frac{1}{\Delta} \left( e^{2 C_1} z_1 \DD z_2 \wedge \DD z_3 + e^{2 C_2} z_2 \DD z_3 \wedge \DD z_1 + e^{2 C_3} z_3 \DD z_1 \wedge \DD z_2\right) -
 \DD z_1 \wedge A_1 
 \nn &-& \DD z_2 \wedge A_2 - \DD z_3 \wedge A_3 - z_1 A_2 \wedge A_3 -z_2 A_3 \wedge A_1 - z_3 A_1 \wedge A_2,  
\eea
incorporates the original transverse NS two-form, $B$, which is unaffected, in addition to generating new terms. The transformed dilaton may be written as  
\begin{equation}
e^{\hat{\Phi}} = e^{\Phi} \Delta^{-\frac{1}{2}}. 
\end{equation}

Using the results of Ref. \cite{Itsios:2013wd}, one can further identity a preferred T-dual frame \footnote{This corresponds to $\hat{e}_-$ in the notation of \cite{Itsios:2013wd}. Switching frames to $\hat{e}_+$ simply involves flipping the sign of $z_a$ in the generated solution. This action does not affect the metric, but changes the sign of the generated NS two-form.} and check that it squares to recover the T-dual metric, 
\bea
\label{Tdual_frame}
\hat{e}^{a}_- = e^{C_a} \Delta^{-1} \biggl[ z_a z_c \DD z_c + e^{2 \sum_{b\neq a} C_b} \DD z_a +\epsilon_{abc} z_b e^{2 C_b} \DD z_c \biggr]. 
\eea
We will henceforth drop subscripts on the frame. 

At this stage, we could isolate a radial direction, $r$, by defining $z_a  = r \mu_a$, where $\mu_a$ would denote constrained scalars on an S$^2$. This approach was adopted in \cite{Jeong:2013jfc}, where it led to a natural SU(2) gauging of the S$^2$. Here we will not consider the rewriting and will choose to retain the $z_a$ as T-dual coordinates. 

It is timely to record expressions for the corresponding derivatives
\bea
\label{derivative}
\hat{\partial}_a &=& \left[  e^{C_a} \partial_{z_a} + \e_{abc} e^{- C_a} z_b \partial_{z_c} \right], \nn
\hat{\partial}_{\mu} &=& \partial_{\mu} - \epsilon_{abc} A^a_{\mu} z_b \partial_{z_c}- B^{a}_{\mu} \partial_{z_a}, 
\eea
which allows us to rewrite $\DD z_a$ in terms of the frame,  
\bea
\DD z_a &=& e^{C_a} \hat{e}^a -\epsilon_{abc} z_b e^{-C_c} \hat{e}^c. 
\eea

\section{Supersymmetry: IIA to IIB}
\label{sec:IIAIIB}
\setcounter{equation}{0}

In this section, we will address supersymmetry in the context of an SU(2) non-Abelian T-duality by studying the transformation of the supersymmetry variations. As we shall see in due course, the variations before and after T-duality are the same provided the Killing spinors are independent of the coordinates of the SU(2) factor on which we dualise. This condition is equivalent to the vanishing of the  Kosmann derivative, an observation that reduces the problem of preserved supersymmetry to either an explicit determination of the Killing spinors or an evaluation of the Kosmann derivative. 

By analogy with the Abelian case, for example \cite{Hassan:1999bv}, or the analysis of section \ref{sec:abelian}, it is expected that the Majorana-Weyl Killing spinors transform as follows
\be
\label{rotation}
\tilde{\epsilon}_- = \epsilon_-, \quad \tilde{\epsilon}_+ = \beta \,  \Omega^{-1} \epsilon_+, 
\ee
where $\beta = \pm 1$ allows for an ambiguity in signs, which will be fixed by combing the NS sector with the RR sector, and $\Omega$ takes the form \cite{{Itsios:2013wd}}
\be
\label{omega}
\Omega = \frac{1}{\sqrt{\Delta}} \left( - \alpha e^{C_1 + C_2 + C_3} \Gamma^{123} + z_a e^{C_a} \Gamma^a \right).  
\ee
Again $\alpha = \pm 1$ allows for a choice of sign, which will get fixed by the T-dual NS two-form (\ref{B1}). Since $\Omega$ has an odd number of gamma matrices, it is clear that it flips the chirality of the spinors. We will adopt $ \G^{0123456789} \tilde{\eta} = - \tilde{\eta}$ in type IIB. 

Now our strategy is morally the same as \cite{Hassan:1999bv}. We will use our knowledge of the transformed NS sector as a means to generate the transformation of the RR fluxes. We will do this by initially comparing the gravitino variation along the transverse seven-dimensional spacetime, $\delta \Psi_{\mu}$. We will then check that the resulting fluxes, when plugged back into the remaining variations, either produce Kosmann derivative terms (\ref{P0}), (\ref{P1}) or the supersymmetry variations of the original spacetime Ansatz. As a further consistency check, we will confirm that the Bianchi identities (appendix \ref{sec:Bianchi}) before and after T-duality agree. One could go further and show that the EOMs match, but it is a given that supersymmetry usually implies the EOMs through integrability and we leave a more elegant proof of this to future work \footnote{A partial direct proof using consistent dimensional reduction can be found in \cite{Jeong:2013jfc}, but all $C_a$ are equal, i. e. there is a round S$^3$, and the flux Ansatz is not completely general.}. 

We have used the SU(2) T-duality rules (\ref{Buscher}) to generate the NS sector. Before T-duality, the gravitino variation may be expressed as 
\bea
 \delta \Psi_{\mu \, \pm } &=& \nabla_{\mu}  \e_{\pm} - \frac{1}{4} \epsilon_{abc} A^{a}_{\mu} e^{C_b-C_c} \G^{bc} \e_{\pm} + \frac{1}{4} e^{C_a} F^a_{\mu \nu} \G^{\nu a} \e_{\pm} \mp \frac{1}{8} H_{\mu \nu \rho} \G^{\nu \rho} \epsilon_{\pm} \mp \frac{1}{4} \DD_{\mu} B_{\rho}^a e^{-C_a} \Gamma^{\mu a} \epsilon_{\pm} \nn &\pm& \frac{1}{8} \epsilon_{abc} B^a e^{-C_b - C_c} \Gamma^{bc} \epsilon_{\pm} +  \frac{1}{8} e^{\Phi} \left[ F_0 \pm \frac{1}{2} \slashed{F}_2  + \frac{1}{24} \slashed{F}_4  \right] \Gamma_{\mu} \epsilon_{\mp}.
\eea
In deriving this expression, we have made use of the spin connection (\ref{spin}) and the field strength $H_3$ (\ref{H0}). So far we have not specified an Ansatz for the RR sector, so $F_p, p =0, 2, 4$ appear. 

Repeating the analysis with the T-dual NS sector, given by expressions (\ref{H1}) and a spin connection that can be derived from (\ref{spin1}), after various cancellations, the T-dual gravitino variation may be massaged into the form
\bea
\delta \hat{\Psi}_{\mu \, -} &=& \nabla_{\mu} {\epsilon}_-   - \frac{1}{4} \epsilon_{abc} A^{a}_{\mu} e^{C_b-C_c} \G^{bc} {\e}_{-} + \frac{1}{4} e^{C_a} F^a_{\mu \nu} \G^{\nu a} {\epsilon}_- +  \frac{1}{8} H_{\mu \nu \rho} \G^{\nu \rho} {\epsilon}_{-} \nn
&+& \frac{1}{4} \DD_{\mu} B_{\rho}^a e^{-C_a} \Gamma^{\mu a} {\epsilon}_{-} -  \frac{1}{8} \epsilon_{abc} B^a e^{-C_b - C_c} \Gamma^{bc} {\epsilon}_{-}, 
\eea
and 
\bea
\beta \, \Omega \, {\delta}  \hat{\Psi}_{\mu +} &=& \biggl( \nabla_{\mu}  {\epsilon}_{+} - \frac{1}{4} \epsilon_{abc} A^{a}_{\mu} e^{C_b-C_c} \G^{bc} {\e}_{+} + \frac{1}{4} e^{C_a} F^a_{\mu \nu} \G^{\nu a} {\epsilon}_+  - \frac{1}{8} H_{\mu \nu \rho} \G^{\nu \rho} {\epsilon}_{+} \nn 
&-& \frac{1}{4} \DD_{\mu} B_{\rho}^a e^{-C_a} \Gamma^{\mu a} {\epsilon}_{+} + \frac{1}{8} \epsilon_{abc} B^a e^{-C_b - C_c} \Gamma^{bc} 
 \epsilon_+ \biggr), 
\eea
where we have re-expressed the T-dual spinor, $\tilde{\eta}$ in terms of the original spinor $\eta$. We observe that the transverse gravitino variation, when restricted to the NS sector, agrees precisely before and after T-duality, providing reassuring confirmation that the SU(2) T-duality rules have been applied correctly. The choice of $\alpha$ in (\ref{omega}) is determined by the sign of the NS two-form (\ref{B1}) and gets fixed to $\alpha =+1$. 

Now that we have some confidence that the NS sector has been T-dualised correctly, we can, given an Ansatz for the original RR sector, generate the T-dual RR sector simply by coercing the supersymmetry variations to match. We are thus free to consider a general Ansatz for the RR sector, one with SU(2) symmetry: 
\bea
\label{IIAfluxes}
F_0 &=& m, \nn
F_2 &=& G_2 + J^{a}_1\wedge (\tau_a + A^a) + \frac{1}{2} \epsilon_{abc} K^a_0 (\tau_b + A^b) \wedge (\tau_c + A^c), \nn
F_4 &=& G_4 + L^a_3 \wedge (\tau_a + A^a) + \frac{1}{2} \epsilon_{abc} M^{a}_2 \wedge (\tau_b + A^b) \wedge  (\tau_c + A^c)  \nn 
&+& N_1 \wedge (\tau_1 + A^1) \wedge (\tau_2 + A^2) \wedge (\tau_3 + A^3),  
\eea
where repeated indices are summed and we have expanded in left-invariant one-forms, which appear wedged with forms living on the transverse seven-dimensional spacetime,  $G_{p}, J_1, K_0, L_3, M_2$ and $N_1$. $m$ is a constant corresponding to the Romans' mass. In the context of non-Abelian T-duality, an RR flux Ansatz comprising the $F_4$ term expanded in (right-invariant) one-forms appeared previously in Ref. \cite{Jeong:2013jfc}, but a thorough supersymmetry analysis was not presented. Here we generalise the Ansatz of Ref. \cite{Jeong:2013jfc} to include an expansion for $F_2$.  Setting all fields except $m, G_2$ and $G_4$ to zero, one recovers the SO(4)-invariant Ansatz of Ref. \cite{Itsios:2012dc}. 

We can now generate the correct T-dual for the RR sector: 
\bea
\label{IIBfluxes}
\hat{F}_1 &=& m \, z_a \, e^{C_a} \hat{e}^a - z_{a} \, J^a_1-K_0^a \, e^{C_a} \hat{e}^a + \e_{abc} K_{0}^a \, z_b \, e^{-C_c} \hat{e}^c  + N_1, \nn
\hat{F}_3 &=& m \, e^{\sum_{a} C_a} \hat{e}^{123} + e^{\sum C_a } *_7 G_4  + G_2 \wedge z_a \, e^{C_a} \hat{e}^a - \frac{1}{2} \e_{abc} J^a_1 \wedge e^{C_b + C_c} \, \hat{e}^{bc}  \nn &+& J^a_1 \wedge e^{-C_a} \hat{e}^a \wedge z_b \, e^{C_b} \hat{e}^b  + z_a \, K_0^a \, e^{2 C_a}  \, e^{-\sum_{b} C_b} \hat{e}^{123}- N_1 \wedge \frac{1}{2} \epsilon_{abc} z_a e^{-C_b-C_c} \hat{e}^{bc} \nn
&-& z_a L^a_3 - M_2^{a} \wedge e^{C_a} \hat{e}^a + \epsilon_{abc} M^a_2 z_b \wedge e^{-C_c} \hat{e}^c,  \nn
\hat{F}_5 &=& (1 + *_{10} ) \biggl[ G_4 \wedge z_a e^{C_a} \hat{e}^a  + e^{\sum C_a} G_2 \wedge \hat{e}^{123}
- \frac{1}{2} \epsilon_{abc} L_3^a \wedge e^{C_b+C_c} \hat{e}^{bc} \nn &+& L_3^{a} \wedge e^{-C_a} \hat{e}^a \wedge z_b \, e^{C_b} \hat{e}^b  + z_a M^a_{2} e^{2 C_a} \wedge e^{-\sum_b C_b} \hat{e}^{123}\biggr], 
\eea
where again repeated indices are summed and we have set $\beta =-1$ for consistency. One could alternatively generate the RR sector as initially suggested in \cite{Sfetsos:2010uq} by acting on the flux bi-spinor with $\Omega^{-1}$, however one is still left the task of confirming the EOMs and the supersymmetry variations and matching the supersymmetry variations directly constitutes an advantageous short-cut.

We now check that our choice of RR sector is consistent with the remaining supersymmetry variations, starting with the gravitino variations along the T-dual directions. Before proceeding, we will explicitly write out an expression for $P_1$. To do this, we recall (\ref{P1}) and insert in the explicit form of $H_3$ from (\ref{H0}) and $F_2, F_4$ from (\ref{IIAfluxes}). This leads to the following lengthy expression: 
\bea
\label{P1_explicit}
P_1 \eta &=& \frac{1}{8} e^{C_1} \slashed{F}^1 \eta + \frac{1}{4} e^{-C_1 -C_2 -C_3} ( e^{2 C_1} - e^{2 C_2} - e^{2 C_3}) \G^{23} \eta - \frac{1}{4} e^{-C_1 -C_3} (e^{2 C_1} - e^{2 C_3}) \slashed{A}^2 \G^{3} \eta \nn
&+& \frac{1}{4} (e^{2 C_1} - e^{2 C_2} ) e^{-C_1 - C_2} \slashed{A}^3 \G^{ 2} \eta + \frac{1}{2} \slashed{\partial} C_1 \G^{1} \eta 
+ \frac{1}{8} \slashed{D} B^1 e^{-C_1} \sigma^3 \eta \nn &+& \frac{1}{4} e^{-C_1 -C_2} \slashed{B}^3 \G^{2} \sigma^3 \eta 
- \frac{1}{4} e^{-C_1 -C_3} \slashed{B}^2 \G^{3} \sigma^3 \eta - \frac{1}{8} e^{\Phi} \biggl[ m \sigma^1 + \frac{1}{2} \slashed{G}_2 (i \sigma^2) \nn &+& \slashed{J}_1^a e^{-C_a} \G^a (i \sigma^2) + \frac{1}{2} \epsilon_{abc} K_0^a e^{-C_b-C_c} \G^{bc} (i \sigma^2) + \frac{1}{24} \slashed{G}_4 \sigma^1 + \frac{1}{6} \slashed{L}_3^a e^{-C_a} \G^{a} \sigma^1  \nn &+& \frac{1}{4}  \epsilon_{abc} \slashed{M}_2^a e^{-C_b - C_c} \G^{bc} \sigma^1 + \slashed{N}_1 e^{-\sum_{a} C_a} \G^{123} \biggr] \G_1 \eta, 
\eea
where repeated indices are assumed to be summed.

After T-duality, we find that the gravitino variations in the T-dual directions can be written in terms of $P_a \eta$: 
\bea
\label{Psi1}
\delta \hat{\Psi}_{1-} &=& \hat{\partial}_{1} {\epsilon}_{-} - P_1 {\epsilon}_{-} + \frac{2}{\Delta} e^{C_1} \biggl[  ( e^{2 C_2 + 2 C_3} + z_1^2) e^{C_1} P_1 {\epsilon}_{-}\nn 
&-&  (e^{2 C_3} z_3 - z_1 z_2) e^{C_2} P_2 {\epsilon}_{-} +  (e^{2 C_2} z_2 + z_1 z_3 ) e^{C_3} P_3 {\epsilon}_{-} \biggr], \nn
\beta \, \Omega \,  \delta \hat{\Psi}_{1+} &=& \hat{\partial}_{1} \e_{+} - P_1 \e_+. 
\eea

As expected we see that the gravitino variation along a representative T-dual direction vanishes provided the Kosmann derivative is zero, $\mathcal{L}_{K_a} \eta =0$, and the derivative with respect to the T-dual directions (\ref{derivative}), $\partial_{z_a}$,  is also zero. Since our Killing spinors did not initially depend on these directions and we have introduced them in the process of performing the T-duality, this latter condition is trivially satisfied. Observe also that due to the cyclic symmetry, it is enough to check one of the internal directions. 

We finally come to the dilatino variation. A similar calculation to above reveals that 
\bea
\label{dilatino_minus}
\delta \hat{\lambda}_- &=& \delta \lambda_- \nn 
&+& \frac{1}{\Delta} \biggl[ (e^{2 C_2 + 2 C_3} + z_1^2) e^{C_1} \G^1 + (e^{2 C_3} z_3 + z_1 z_2) e^{C_2} \G^2 - (e^{2 C_2} z_2 - z_1 z_3) e^{C_3} \G^3 \biggr]   e^{C_1} P_1 \epsilon_{-} \nn
&+& \frac{1}{\Delta} \biggl[ (e^{2 C_3 + 2 C_1} + z_2^2) e^{C_2} \G^2 + (e^{2 C_1} z_1 + z_2 z_3) e^{C_3} \G^3 - (e^{2 C_3} z_3 - z_1 z_2) e^{C_1} \G^1 \biggr]   e^{C_2} P_2 \epsilon_{-} \nn
&+& \frac{1}{\Delta} \biggl[ (e^{2 C_1 + 2 C_2} + z_3^2) e^{C_3} \G^3 + (e^{2 C_2} z_2 + z_1 z_3) e^{C_1} \G^1 - (e^{2 C_1} z_1 - z_2 z_3) e^{C_2} \G^2 \biggr]   e^{C_3} P_3 \epsilon_{-} \nn
&=& \delta \lambda_- + \sum_{a=1}^3 \frac{1}{2} \G^{a} \left[ \delta \hat{\Psi}_{a-} + P_{a} \epsilon_{-} \right], 
\eea
where we have made use of (\ref{Psi1}), various cyclic expressions and $\hat{\partial}_a \eta = 0$ to rewrite the above expression. Similarly, one can express $\delta \hat{\lambda}_+$ in terms of $P_a \eta$:
\bea
\beta \, \Omega \, \delta  \hat{\lambda}_{+} &=& \delta \lambda_+ \nn
&-& \frac{1}{\Delta} \biggl[ (e^{2 C_2 + 2 C_3} + z_1^2) e^{C_1} \G^1 - (e^{2 C_3} z_3 - z_1 z_2) e^{C_2} \G^2 + (e^{2 C_2} z_2 + z_1 z_3) e^{C_3} \G^3 \biggr]   e^{C_1} P_1 \epsilon_{+} \nn
&-& \frac{1}{\Delta} \biggl[ (e^{2 C_3 + 2 C_1} + z_2^2) e^{C_2} \G^2 - (e^{2 C_1} z_1 - z_2 z_3) e^{C_3} \G^3 +(e^{2 C_3} +z_3  z_1 z_2) e^{C_1} \G^1 \biggr]   e^{C_2} P_2 \epsilon_{+} \nn
&-& \frac{1}{\Delta} \biggl[ (e^{2 C_1 + 2 C_2} + z_3^2) e^{C_3} \G^3 - (e^{2 C_2} z_2 -z_1 z_3) e^{C_1} \G^1 + (e^{2 C_1} z_1 + z_2 z_3) e^{C_2} \G^2 \biggr]   e^{C_3} P_3 \epsilon_{+}.\nn
\eea
Note that up to a change in sign in terms even under $z_{a} \leftrightarrow -z_a$, the above expression is the same as (\ref{dilatino_minus}). 

This brings our rewriting of the supersymmetry variations to an end. In all equations we have noted that the Killing spinor equations after an SU(2) non-Abelian T-duality can be written in terms of the Killing spinor equations before T-duality up to the vanishing of $P_a \eta$. As discussed in the previous section, $P_a \eta = 0$ is equivalent to the vanishing of the Kosmann derivative with respect to the (right) SU(2) isometry group. This condition further implies that the Killing spinors are constant with respect to these directions. 

Therefore the problem of establishing the extent of supersymmetry breaking under non-Abelian T-duality boils down to an analysis of the Killing spinors themselves. One can either break supersymmetry by imposing suitable projection conditions, thus imposing that the Killing spinors are constant with respect to the T-duality directions, or one can analyse the Kosmann derivative to extract projection conditions.  This observation explains why supersymmetry is broken in the background AdS$_5 \times$ S$^5$ under an SO(6) transformation \cite{Lozano:2011kb}; there are no Killing spinors that are constant with respect to the action of the full SO(6) isometry.

\section{IIB to IIA} 
\label{sec:IIBIIA}
\setcounter{equation}{0}

In this section we repeat the process of the section \ref{sec:IIAIIB}, except this time describing the transformation from type IIB to type IIA supergravity. More concretely, we will adopt the same results for the NS sector, impose the same Ansatz for the Killing spinor transformation (\ref{rotation}) and use supersymmetry to extract the transformed RR fluxes given an initial flux Ansatz. We will assume that there is no deviation from the analysis of the previous section and that one can indeed show that transformed Killing spinor equations correspond to the original Killing spinor equations modulo the appearance of the Kosmann derivative. 

We consider a general RR flux Ansatz with a right-acting SU(2) symmetry. In particular, we write this as \footnote{The inclusion of an SU(2) one-form in $F_1$ is inconsistent with the Bianchi $\dd F_1 = 0$. }
\bea
\label{original IIB fluxes}
F_1 &=& G_1, \nn
F_3 &=& G_3 + X^a_2 \wedge (\tau_a +A^a) +\frac{1}{2}   \epsilon_{abc} Y_{1}^{a} \wedge (\tau_b + A^b) \wedge (\tau_c + A^c) \nn 
&+& m (\tau_1 + A^1) \wedge (\tau_2+A^2) \wedge (\tau_3 + A^3),  \nn
F_5 &=& (1+ *_{10}) \biggl[ Z^{a}_4 \wedge (\tau_a+ A^a) + G_2 \wedge (\tau_1 + A^1) \wedge (\tau_2 + A^2) \wedge (\tau_3+ A^3)   \biggr],
\eea
where $m$ is a suggestive constant and we have used $G_p$, $p=1, 2, 3$, to label the SO(4)-singlets that appeared in Ref. \cite{Itsios:2012dc}. The remaining fields,  $ X_2, Y_1$ and $Z_4$  are novel to the analysis presented in this paper. Owing to self-duality of the five-form flux, terms involving an even number of products of the base one-form $(\tau_a + A^a)$ are implied. There are numerous explicit examples of SU(2) transformations covered by our Ansatz in the literature \cite{Sfetsos:2010uq, Itsios:2013wd,  Sfetsos:2014tza, Macpherson:2013zba, Gaillard:2013vsa}. 

From an analysis of the gravitino variation, $\delta \Psi_{\mu}$, one can read off the T-dual RR flux: 
\bea\label{IIA fluxes}
\hat{F}_0 &=& -m, \nn
\hat{F}_2 &=& z_a e^{C_a} \hat{e}^a \wedge G_1  + z_a X_2^a + Y_1^a \wedge e^{C_a} \hat{e}^a - \epsilon_{abc} Y_1^a \wedge z_b e^{-C_c} \hat{e}_c + m \frac{1}{2} \epsilon_{abc} z_a e^{-C_b -C_c} \hat{e}^{bc}, \nn
&-& G_2,  \nn
\hat{F}_4 &=& - e^{\sum_b C_b} G_1 \wedge \hat{e}^{123} + z_a e^{C_a} \hat{e}^a \wedge G_3  + e^{\sum_b C_b} *_7 G_3  + \frac{1}{2} \epsilon_{abc} X_2^{a} \wedge e^{C_b + C_c} \hat{e}^{bc} \nn
&-& X_2^a e^{-C_a} \hat{e}^a \wedge z_b e^{C_b} \hat{e}^b -z_a e^{2 C_a} Y_1^a \wedge e^{-\sum_b C_b} \hat{e}^{123} - z_a Z_4^{a} - e^{\sum_b C_b} *_7  Z^a_4 \wedge  e^{-C_a} \hat{e}_a \nn &+& \epsilon_{abc} e^{-C_a} *_7 Z_4^a \wedge z_b e^{C_b} \hat{e}^c 
+ G_2 \wedge \frac{1}{2} \epsilon_{abc}  z_a e^{-C_b-C_c} \hat{e}^{bc}. 
\eea

As expected, $m$ clearly corresponds to the Romans' mass. Combined with earlier results in this paper, this gives a complete mapping for type IIB supergravity solutions with SU(2) isometry. 

\section{Examples} 
The results of the previous sections provide a comprehensive proof that the question of supersymmetry preservation under U(1) and SU(2) T-duality boils down to checking the Kosmann derivative along the given isometry in the original solution. This is a powerful result with great utility and in this section we will illustrate this point with some recent and new examples. 

We begin by showing how supersymmetry is preserved in recent solutions generated by non-Abelian T-duality acting on AdS$_5~ \times$ Y$^{p,q}$ and AdS$_4\times \mathbb{C P}^3$ \cite{Lozano:2014ata,Sfetsos:2014tza}. We then generate a new AdS$_3$ solution in type IIB  preserving 16 supercharges and leave a discussion of further examples for concluding remarks. In the explicit examples we study we do not present the form of the dual Killing spinor, as in each case, this is defined in terms of the spinor of the original geometry (\ref{rotation}).

Let us start with a recap of the main result; there are essentially two cases:
\begin{enumerate}
\item Supersymmetry is preserved when the Kosmann derivative of the original Killing spinor $\eta$ w. r. t the T-duality directions vanishes.
\item Supersymmetry is broken whenever the Kosmann derivative w. r. t. the T-duality directions is non-zero, with the degree of supersymmetry breaking determined by the Kosmann derivative.  
\end{enumerate}
As we have elucidated earlier, when the Kosmann derivative vanishes, the Killing spinors are independent of the isometry directions making it intuitively obvious that supersymmetry will be preserved. Therefore, where the original Killing spinors are known, or the initial geometry is suitably simple, it may be easier to directly examine the Killing spinors. Whenever this proves problematic, one still has recourse to the Kosmann derivative. The latter is of course  frame independent, however it is useful to work in the preferred frame of T-duality of (\ref{eq: TDframe}) or (\ref{eq: NATDframe}), since after all, one needs to be in this frame to read off the T-dual solution. This reduces to applying equations (\ref{B1}), (\ref{Tdual_frame}) and either (\ref{IIBfluxes}) or (\ref{IIA fluxes}). 


To illustrate our methods, let us first look at a simple case that preserves all the supercharges of the original solution, namely the SU(2) transformation of AdS$_5~ \times$ Y$^{p,q}$. This example was recently presented in \cite{Sfetsos:2014tza}. 

\subsection*{The Dual of AdS$_5 ~\times$ Y$^{p,q}$}
We recall that Y$^{p,q}$ is an infinite family \cite{Gauntlett:2004yd} of Sasaki-Einstein manifolds that, when embedded in type-IIB supergravity, is dual to $\mathcal{N}=1$ superconformal quivers in four dimensions \cite{Martelli:2004wu,Benvenuti:2004dy}. They can be realised as the near-horizon limit of of a stack of $N$ D3 branes placed at the tip of the associated Calabi-Yau cone. 

In \cite{Sfetsos:2014tza} this geometry was T-dualised along its SU(2) isometry and it was shown that $\mathcal{N}=1$ supersymmetry was preserved by demonstrating that an SU(2)-structure exists on the internal space \footnote{See \cite{Grana:2005sn} for the original work on G-structures in the presence of non trivial RR fields.}. It was additionally shown that the Kosmann derivative vanishes, implying directly through our earlier work that supersymmetry is preserved. For completeness, here we use results in the literature, e.g. \cite{Canoura:2005uz}, to show that the Killing spinor in the frame of equation (\ref{eq: NATDframe}) is independent of the isometry directions. 

The Killing spinors of AdS$_5~\times$ Y$^{p,q}$ were originally derived in \cite{Canoura:2005uz} in the orthonormal frame ($c=1$)
\begin{align}\label{eq: ypqveil1}
&e^{x^{\mu}}= r \dd x^{\mu},~~~ e^{r}=\frac{\dd r}{r},~~~e^{y}=-\frac{\dd y}{\sqrt{wv}},~~~e^{\beta}=-\frac{\sqrt{wv}}{6}(\dd\beta-\cos\theta \dd \phi),\nn
&e^1=\frac{\sqrt{1-y}}{\sqrt{6}}\dd \theta,~~~ e^{2}=\frac{\sqrt{1-y}}{\sqrt{6}}\sin\theta \dd \phi,~~~ e^3=\frac{1}{3}\big(\dd \psi+y \dd \beta+ (1-y) \cos \theta \dd\phi\big),
\end{align}
where we have defined 
\be
w = \frac{2 (a-y^2)}{1-y}, \quad  v = \frac{a - 3 y^2 + 2 y^3}{a-y^2}, 
\ee
with $a$ a constant. The Killing spinor is determined to be
\begin{equation}\label{eq: ypqspinor1}
\eta= e^{-\frac{i}{2}\psi}r^{-\frac{\Gamma_{\star}}{2}}\bigg(1+\frac{\Gamma_r}{2}x^{\mu}\Gamma_{{\mu}}(1+\Gamma_{\star})\bigg)\eta_0
\end{equation}
where $\Gamma_{\star}=i\Gamma_{x^0x^1x^2x^3}$ and $\eta_0$ is a constant spinor satisfying the projection conditions \footnote{Here $\eta$ is a complex Killing spinor in type IIB. The real Majorana-Weyl spinors can be extracted via $\eta = \epsilon_1 + i \epsilon_2$. This change must be supplemented in the projections by the replacement $i\eta\to (i\sigma_2)\eta$.}
\begin{equation}
\Gamma_{y \beta}\eta_0=-i\eta_0,~~~ \Gamma_{12}\eta_0= i \eta_0. 
\end{equation}
We observe that in this frame, the natural frame for Abelian T-duality with respect to the isometry $\partial_{\psi}$, the Killing spinors necessarily depend on $\psi$, so supersymmetry will be completely broken under Abelian T-duality. This is not the case for an SU(2) T-duality, which we will show by exploiting a frame rotation to recast (\ref{eq: ypqveil1}) in terms of a natural frame for SU(2) T-duality \cite{Sfetsos:2014tza}. More precisely, we consider the following rotations \footnote{In the process of performing these rotations we also redefine $\beta = - (6 \alpha +\psi)$ and relabel the frame accordingly.}:
\begin{align}
&e^{\alpha}~\!\!'=\cos\lambda e^{\beta}+\sin\lambda e^{3}=  \frac{1}{3} \frac{\sqrt{wv}}{\sqrt{g }}\dd \alpha,\nn[2mm]
&e^{3}~\!\!'=\cos\lambda e^{3}-\sin\lambda e^{\beta}=(\sqrt{g}\tau_3+\frac{w f}{\sqrt{g}}  \dd \alpha),\nn[2mm]
&e^{1}~\!\!'=\cos\psi e^{1}+\sin\psi e^2=\frac{\sqrt{1-y}}{\sqrt{6}}\tau_1,~~~e^2~\!\!'=\cos\psi e^2-\sin\psi e^{1}=\frac{\sqrt{1-y}}{\sqrt{6}}\tau_2,\nn[2mm]
&e^a~\!\!'=e^a,~~~ a=x^{\mu},r,y, 
\end{align}
where
\begin{equation}
f = \frac{a-2 y + y^2}{6 (a-y^2)}, \quad \cot\lambda =\frac{2 (1-y)}{\sqrt{w v}}, \quad g = \frac{v}{9} + w f^2. 
\end{equation}
Such rotations act on the spinor by a matrix $\mathcal{S}$ which satisfies $\mathcal{S}^{-1}\Gamma_a\mathcal{S}=\Lambda_{a}^{~b} \G_{b}$. The solution for a rotation by $\chi$ acting on flat directions $z_1,z_2$ is simply
\begin{equation}
\mathcal{S}_{\chi}= e^{\frac{1}{2}\chi\Gamma_{z_1z_2}},
\end{equation}
which results in a spinor independent of the SU(2) directions $\theta,\phi,\psi$, namely
\begin{equation}
\eta'=e^{\frac{1}{2}\lambda \Gamma_{\alpha3 }'}r^{-\frac{\Gamma_{\star}'}{2}}\bigg(1+\frac{\Gamma_r'}{2}x^{\mu}\Gamma_{^{\mu}}'(1+\Gamma_{\star}')\bigg)\eta_0
\end{equation}
Since the Killing spinor is independent of the SU(2) directions, it is now obvious that $\mathcal{N}=1$ supersymmetry is preserved under the SU(2) transformation.

Adopting our notation and $(7,3)$-split for the spacetime, the solution may be expressed in the form (\ref{metric}) as
\begin{align}
&\dd s^2_7= \dd s^2(AdS_5)+ \frac{\dd y^2}{w v}+ \frac{w v}{9 g} \dd \alpha^2,\nn
&e^{2C_1}=e^{2C_2}=\frac{1-y}{6},~~~e^{2C_3}=g,\nn
&A^1=A^2=0,~~~A^3=\frac{w f }{g} \dd \alpha. 
\end{align}
Further comparison with (\ref{original IIB fluxes}) means the metric is supported by non trivial RR 5-form flux with
\begin{equation}
G_2 =-\frac{2}{9} (1-y) \dd y\wedge \dd \alpha. 
\end{equation}
All other terms in (\ref{original IIB fluxes}) are zero and the dilaton is a constant. One is now in a position to generate the solution using earlier results and we omit further details. 

\subsection*{The Dual of AdS$_4 \times \mathbb{C P}^3$}
AdS$_4 \times \mathbb{C P}^3$ is a solution in type IIA first introduced in \cite{Nilsson:1984bj} and later identified as the holographic description of $\mathcal{N}=6$ Chern-Simons matter theory \cite{Aharony:2008ug}. The geometry preserves 24 supersymmetries. To see this, one can decompose the general Killing spinor $\eta$ as a product of four and six-dimensional spinors $\eta=\eta_{AdS_4}\otimes\eta_{\mathbb{CP}^3}$, where the standard spinor on AdS$_4$, $\eta_{AdS_4}$ preserves four supercharges. In contrast to the external spinor's simplicity, the internal spinor is generally quite complicated and depends on all coordinates on $\mathbb{CP}^3$. Despite the complexity, it can be shown that $\mathbb{CP}^3$ preserves a 6-component spinor \cite{Nilsson:1984bj}, giving a total of 24 preserved supersymmetries. 

It has recently been shown \cite{Lozano:2014ata} (see appendix B for details) that when the internal spinor has only two components, a frame exists
\begin{align}\label{eq:frameABJM}
& e^{\zeta}= L \dd \zeta,~~~ e^{\theta_1}= \frac{L}{2}\cos\zeta \dd \theta_1,~~~e^{\phi_1} = \frac{L}{2}\cos\zeta\sin\theta_1 \dd \phi_1,\nn
& e^{1} = \frac{L}{2}\sin\zeta\,\tau_1,~~~e^{2} = \frac{L}{2}\sin\zeta\,\tau_2,~~~e^{3}= \frac{L}{2}\sin\zeta\cos\zeta(\tau_3+\cos\theta_1 d\phi_1),
\end{align}
where the truncated Killing spinor depends only on two arbitrary constants $a, b$ and may be expressed as \footnote{Here we use a specific representation of the gamma matrices in 6d,
\begin{equation}
\begin{array}{ccccccccccccccc}
\gamma^{\zeta} &=&1&\otimes& \s^2&\otimes &\s^1,&~~~&\gamma^{\theta_1} &=&1&\otimes& \s^2&\otimes &\s^3,\\
\gamma^{\phi_1} &=&\s^1&\otimes& 1&\otimes& \s^2,&~~~&\gamma^1 &=&\s^3&\otimes& 1&\otimes &\s^2,\\
\gamma^2 &=&\s^2&\otimes& \s^1&\otimes& 1,&~~~&\gamma^3 &=&\s^2&\otimes &\s^3&\otimes& 1,
\end{array}
\end{equation}
where $\sigma^i$ are the Pauli matrices.}
\begin{equation}
\hat{\eta}_{\mathbb{CP}^3}=(0,a \cos\zeta,a\sin\zeta,0,0,-b\sin\zeta,b\cos\zeta,0)^T
\end{equation}
With this choice all dependence on the isometry directions drops out. Thus we find that performing an SU(2) T-duality on AdS$_4 \times \mathbb{CP}^3$ preserves a total of 8 supercharges, resulting in $\mathcal{N}=2$ supersymmetry in three dimensions.\\

The spacetime metric can be rewritten in terms of our Ansatz (\ref{metric}) as
\begin{align}
& \dd s^2_7= \frac{L^2}{4}\dd s^2(AdS_4)+ L^2 \left[ \dd \zeta^2+\frac{1}{4}\cos\zeta^2 ( \dd \theta_1^2+\sin^2 \theta_1 \dd \phi_1^2) \right],\nn
&e^{2C_1}=e^{2C_2}=\frac{L^2}{4}\sin^2\zeta,~~~e^{2C_3}=\frac{L^2}{4}\sin^2 \zeta\cos^2 \zeta,\nn
&A^1=A^2=0,~~~A^3=\cos \theta_1 \dd \phi_1.
\end{align}
The constant dilaton is given by $e^{\Phi}=\frac{L}{k}$ and NS 2-form is gauge trivial. The RR sector, in terms of out notation (\ref{IIAfluxes}), has the following non-zero seven-dimensional components
\begin{align}
& G_2=-\frac{k}{2}\cos^2\zeta \sin \theta_1 \dd \theta_1 \wedge \dd \phi_1,~~~J_1^3=-k \sin\zeta\cos\zeta \dd \zeta,~~~K^1_0=-\frac{kL}{2}\sin^2\zeta,\nn
&G_4=-\frac{3k L^2}{8} \vol(AdS_4).
\end{align}
The SU(2) transformation of this solution was performed in \cite{Lozano:2014ata}, where it was claimed that the generated geometry describes a strongly-coupled three-dimensional quiver preserving $\mathcal{N}=2$ supersymmetry. It was assumed that the Kosmann derivative suitably captured the preserved supersymmetry and in this paper we confirm that the assumption is correct. 


\subsection*{A new AdS$_3$ solution preserving 8 Supercharges}
In this section we generate a new AdS$_3$ solution in type IIA supergravity by performing an SU(2) transformation on a well-known AdS$_3 \times$ S$^3 \times$ S$^3\times \mathbb{R}$ geometry. We consider the solution in type IIB presented in \cite{de Boer:1999rh}, which corresponds to the near-horizon limit of two stacks of D5 branes with D1 branes lying at their intersection and smeared elsewhere \cite{Cowdall:1998bu,Boonstra:1998yu,Gauntlett:1998kc} (see also \cite{Donos:2008hd}). The metric may be expressed as
\begin{equation}
\dd s^2= L^2 \dd s^2(AdS_3)+R_1^2 \dd s^2(S^3_1)+R_2^2 \dd s^2(S^3_2)+\dd x^2, 
\end{equation}
while the dilaton is  constant. The metric is supported by a non-trivial RR three-form
\begin{equation}
F_3= 2 L^2 \vol(AdS_3)+2 R_1^2 \vol(S^3_1)+2 R_2^2 \vol(S^3_2).
\end{equation}
From the EOMs we get the constraint on the radii
\begin{equation}
\frac{1}{L^2}=\frac{1}{R_1^2}+\frac{1}{R_2^2}.
\end{equation}
This solution has attracted some recent interest, particularly so since a proposal for the dual CFT was given in \cite{Tong:2014yna} (see  \cite{de Boer:1999rh,Gukov:2004ym} for earlier work). The dual CFT is a $\mathcal{N}=(4,4)$ gauge theory with two copies of SU(2) $\times$ SU(2) R-symmetry, realised in the geometry by the two round 3-spheres. Since this corresponds to two copies of large superconformal symmetry \cite{Sevrin:1988ew}, we expect that T-dualising with respect to $SU(2) \subset SO(4)$, one will realise a geometry preserving only one copy of the large superconformal symmetry, so $\mathcal{N} = (0,4) $ supersymmetry \footnote{By performing Abelian T-duality on a linear combination of the Hopf-fibres, one can preserve $\mathcal{N} = (4,2)$ supersymmetry \cite{Donos:2014eua}.}.  


We will now use our earlier results to T-dualise with respect to one of the SU(2) isometries of $S^3_2$ and from here on set $S^3_1=S^3$. We read off the following input for the SU(2) rules
\begin{align}
&e^{2C_a}= \frac{R_2^2}{4},~~~~a=1,2,3,\nn
&G_3= 2 L^2 \vol(AdS_3)+2 R_1^2 \vol(S^3),~~~ m = \frac{R_2^2}{4},
\end{align} 
with all other fields zero. We note that the appearance of a non-zero Romans' mass $m$ indicates that the T-dual solution will be a massive type IIA solution. We can immediately read off the T-dual solution, however to make the remnant SU(2) symmetry manifest, it is advantageous to express the T-dual geometry in spherical polar coordinates
\begin{equation}
v_1=\rho\sin\chi\cos\xi,~~~v_2=\rho\sin\chi\sin\xi,~~~v_3=\rho\cos\chi.
\end{equation}
The resulting  metric is
\begin{equation}
ds^2_{IIA}= L^2 \dd s^2(AdS_3)+ \dd x^2+ R_1^2 \dd s^2(S^3)+\frac{4}{R_2^2}\bigg( \dd \rho^2+\frac{ R_2^6 \rho^2}{64\Delta}\dd s^2(S^2)\bigg),
\end{equation}
where $S^2$, corresponding to the SU(2) R-symmetry, is the two-sphere spanned by $\chi$ and $\xi$ and 
\begin{equation}
\Delta= \frac{R_2^6+16 R_2^2\rho^2}{64}.
\end{equation}
The dual dilaton is given by
\begin{equation}
e^{-2\Phi}={\Delta}, 
\end{equation}
while the NS two-form is simply
\begin{equation}
B_2= -\frac{ R_2^2 \rho^3}{4 \Delta} \vol(S^2).
\end{equation}

The RR sector can be read off from eq. (\ref{IIA fluxes}) and is given by
\begin{align}
m&=-\frac{R_2^2}{4},~~~~F_2= m B_2,\\
F_4&= \frac{ R_2^3}{4L R_1}\big[ L^2 \vol(AdS_3)+R_1^2 \vol(S^3)\big]\wedge \dd x + {2 \rho}\big[L^2 \vol(AdS_3)+R_1^2 \vol(S^3)\big]\wedge \dd \rho.\nonumber
\end{align} 
We have investigated the curvature invariants of this solution and found that they are non-singular. This is in line with our expectations, since we dualised on an S$^3$ of constant radius.

\section{Conclusions}
For spacetimes with SO(4) isometry it has been demonstrated that left or right acting SU(2) non-Abelian T-duality is a symmetry of the EOMs of type II supergravity and, in addition, that the supersymmetry conditions before and after T-duality are mapped into each other up to the vanishing of the Kosmann spinorial Lie derivative \cite{Itsios:2012dc}. Since a great number of interesting geometries, for example \cite{Itsios:2013wd, Lozano:2014ata, Sfetsos:2014tza}, fall outside the scope of this earlier work, in this paper we have extended the supersymmetry analysis of Ref. \cite{Itsios:2012dc} to a large class of solutions with Bianchi IX symmetry. In analogy with the Abelian case, we have noted that the Kosmann derivative captures spacetime supersymmetry breaking. Therefore, any question regarding spacetime supersymmetry can be settled by studying the Kosmann derivative, or alternatively, by explicitly working out the Killing spinors of the original geometry. It should be instructive to recast these findings in terms of the language of pure spinors. 

It is expected that an underlying inert seven-dimensional theory can be found via consistent Kaluza-Klein reduction in a similar vein to \cite{Itsios:2012dc}. This would provide an effective, yet formal proof that the SU(2) transformation is also a symmetry of the EOMs for the general Ansatz we consider. Evidence suggests that this theory is not a supergravity and will possess massive modes \cite{Itsios:2012dc}. Given that supersymmetry under mild assumptions implies certain EOMs, it may be more elegant to provide a proof based on supersymmetry, and we leave this to future work. 

At this stage it is firmly believed that non-Abelian T-duality takes supergravity solutions to supergravity solutions. However, there is currently no definitive statement when $\alpha'$-corrections to supergravity are considered. This motivates the future study of non-Abelian T-duality in the context of Heterotic supergravity, where $\alpha'$ corrections appear at the linear level, to determine whether non-Abelian T-duality is simply a symmetry of the strict supergravity limit, or not. One may eventually consider extending this avenue of enquiry to higher-order corrections, a direction of research that has been pursued in the Abelian case in, for instance, \cite{Meissner:1991zj, Sen:1991zi}. 

Indeed, the form of the transformation (\ref{omega}) is already deeply suggestive, since in various limits, for example $z_1 \rightarrow \infty$, $z_2, z_3$ finite, we recover the results of Abelian T-duality. Thus, by treating the original and T-dual coordinates on an equal footing, as is the case in Double Field Theory (DFT) \cite{Siegel:1993xq, Hull:2009mi}, it should be possible to combine Abelian and non-Abelian T-duality. Such a generalisation already exists in the literature and goes under the name Poisson-Lie T-duality \cite{Klimcik:1995ux, Klimcik:1995jn}, where a natural counterpart of O(d,d) symmetry involves automorphisms of the Drinfeld double, the algebraic structure underpinning Poisson-Lie T-duality. An analogue of DFT for non-Abelian T-duality promises to extend the known class of non-geometric compactifications and preliminary work in this direction is currently underway \cite{Plauschinn:2014nha}. 

Given that non-Abelian T-duality is expected to preserve integrability, another exciting open direction is to consider using it as a means to generate integrable deformations of AdS/CFT geometries.  Given backgrounds with U(1) $\times$ U(1) isometry, it is a well-known fact that integrable deformations based on TsT transformations exist \cite{Lunin:2005jy}. The counterpart in our setting will involve some mixing between residual isometries and a necessary inverse non-Abelian T-duality, in contrast to the usual Abelian case. More generally, there is an open question about how non-Abelian T-duality combines with SL(2,$\mathbb{R}$) (S-duality) transformations in the supergravity and whether this leads to new examples of compact geometries. In this sense, the role of non-Abelian T-duality in solution generating, namely as an intermediary step, may mirror fermionic T-duality \cite{fTduality}. We hope to report on this in future work. 

There is also an ongoing program of work where non-Abelian T-dual geometries are being studied in the context of the AdS/CFT correspondence in order to infer properties of putative dual CFTs \cite{ Itsios:2012zv, Itsios:2013wd, Lozano:2014ata, Sfetsos:2014tza}. Certain examples have been proved to preserve $\mathcal{N}=1$ in four dimensions by an explicit illustration of the dual $G$-structures defined on the six-dimensional internal space \cite{Barranco:2013fza, Gaillard:2013vsa}. Although such an analysis provides additional geometric information about these solutions, the results here are not confined by dimensionality and make any question of supersymmetry rather trivial in these examples. All these solutions are independent of the SU(2) directions in the preferred frame \cite{Arean:2006nc}, and so supersymmetry is preserved. 

There are several novel examples that have yet to be studied, but we can be confident supersymmetry will be preserved under the SU(2) transformation. Examples include the holographic duals of flows from $\mathcal{N} =4$ super-Yang-Mills \cite{Khavaev:1998fb, Pilch:2000ej, Pilch:2000ue}. Indeed, non-Abelian T-duality on the $\mathcal{N}=1$ fixed-point solution \cite{Pilch:2000ej} provides a natural generalisation of the recent work of Ref. \cite{Sfetsos:2014tza} for five-manifolds not in the Sasaki-Einstein class. One might go further and consider non-Abelian T-dualising the known flows between S$^5$ and T$^{1,1}$ \cite{Halmagyi:2005pn,Klebanov:2007us} with the hope of learning something about flows between the $\mathcal{N}=1$ and $\mathcal{N}=2$ Sicilian quivers \footnote{These were argued to be related to the non-Abelian T-dual of T$^{1,1}$  and S$^5$ respectively in \cite{Itsios:2013wd} and \cite{Sfetsos:2010uq}.}. Owing to the fact that the T-dual geometry and preserved supersymmetry can be simply read off from our results, there are now numerous other possibilities to consider.

\section*{Acknowledgements} 
We thank Koushik Balasubramanian, Tom\'as Ort\'in, Martin Ro\v{c}ek, Daniel Thompson and Rikard von Unge for correspondence and discussion. We are grateful to Martin Ro\v{c}ek and Kostas Sfetsos for readings of later drafts and sharing their constructive comments. We thank Thiago Araujo for pointing out an inconsistency in signs in section 6.  O. K. \& E. \'O C. wish to acknowledge support from the Simons Center for Geometry and Physics during the 2014 Simons Summer workshop in Mathematics and Physics.  Y.L. is partially supported by the Spanish Ministry of Science and Education grant FPA2012-35043-C02-02. 
N.T.M is supported by an STFC studentship. The research of Y.L. and N.T.M. has been supported by the EU-COST Action MP1210 ``The String Theory Universe''. E. \'O C. is supported by a Marie Sklodowska-Curie Fellowship, ``T-Dualities". 

\appendix 

\section{Bianchi IX symmetry}
\label{sec:squash} 

\setcounter{equation}{0}
In this paper we focus on internal spaces with SU(2) isometry corresponding to diagonal Bianchi IX symmetry. In terms of left-invariant one-forms, $\tau_a$, these spaces may be written as  
\be
\label{diagSU2_metric}
\dd s^2_{10} = \dd s_7^2 + \sum_{a=1}^3 e^{2 C_{a}} (\tau_a + A^a)^2, 
\ee
where $C_{a}$ and $A^a$ depend on transverse directions. When the scalars $C_a$ are completely generic, the metric has a right-acting SU(2) isometry. When two of the $C_{a}$ are the same, we find the enhancement to SU(2) $\times$ U(1) isometry and when all $C_{a}$ are the same, we recover the round S$^3$. In the last case,  we have SO(4) $\simeq$ SU(2) $\times$ SU(2) isometry and  the Killing vectors can be divided into right-invariant 
\bea
\label{right_inv}
K_1 &=& - \frac{\cos \phi}{\sin \theta} \partial_{\psi} + \sin \phi \partial_{\theta} + \cot \theta \cos \phi \partial_{\phi}, \nn
K_2 &=& \frac{\sin \phi}{\sin \theta} \partial_{\psi} + \cos \phi \partial_{\theta} - \cot \theta \sin \phi \partial_{\phi}, \nn
K_3 &=& \partial_{\phi}, 
\eea
and left-invariant vectors, which are dual to the one-forms appearing in the metric (\ref{diagSU2_metric}), 
\bea
\label{left_inv}
\tilde{K}_1 &=& - \cot \theta \cos \psi \partial_{\psi} - \sin \psi \partial_{\theta} + \frac{\cos \psi}{\sin \theta} \partial_{\phi}, \nn 
\tilde{K}_2 &=& - \cot \theta \sin \psi \partial_{\psi} +\cos \psi \partial_{\theta} + \frac{\sin \psi}{\sin \theta} \partial_{\phi}, \nn 
\tilde{K}_3 &=& \partial_{\psi}. 
\eea

For the metric to be invariant under these symmetries, we require that 
\be
\mathcal{L}_{X} g_{\mu \nu} = X^{\xi} \partial_{\xi} g_{\mu \nu} + \partial_{\mu} X^{\lambda} g_{\lambda \nu} + \partial_{\nu} X^{\lambda} g_{\mu \lambda} = 0, 
\ee
where $X$ is a vector field. It is easy to check that when the $C_{a}$ are generic, the isometries are broken to SU(2) corresponding to the right-invariant vectors, $K_a$. 

Requiring that the Ansatz for the fluxes has the same SU(2) isometry with respect to the right-invariant vectors, fixes them to be written in terms of wedge products of the left-invariant one-forms as presented in the body of the text. Again, a short calculation reveals that for $\omega = (\tau_a + A^a)$ 
\be
\mathcal{L}_{K_{a}} \omega = i_{K_a} \dd \omega + \dd i_{K_a} \omega = 0, 
\ee
where $K_a$ denotes right-invariant vector fields. 

We remark that a more general metric with a right-acting SU(2) isometry takes the form 
\be
\label{genSU2_metric}
\dd s^2 = \sum_{a,b=1}^3 e^{2 C_{ab}} (\tau_a + A^a) \otimes (\tau_b + A^b).  
\ee
While one can always diagonalise the metric at a given point of the transverse space, in general the metric has off-diagonal components. 

\section{Technical Details} 
\setcounter{equation}{0}
In this section we house information about the T-dual field strength $\hat{H}_3$ for the NS two-form and the T-dual spin connection, $\hat{\omega}$,  both of which appear in the T-dual Killing spinor equations. 

In our chosen frame (\ref{Tdual_frame}), the field strength, $\hat{H}_3 = \dd \hat{B}_2$, may then be written as 
\bea
\label{H1}
\hat{H}_3 &=& {-}e^{-C_1 - C_2 -C_3} \left[ \sum_{a} e^{2 C_a} - \frac{2}{\Delta} (\sum_a e^{4 C_a} z_a^2) \right] \hat{e}^{123} \nn
&+& \frac{2}{\Delta} e^{C_1 + C_3} \left[  \dd C_1 (e^{2 C_2} z_2 {-} z_1 z_3)+ \dd C_3 (e^{2 C_2} z_2 {+}z_1 z_3)\right] \hat{e}^{31} \nn
&+& \frac{2}{\Delta} e^{C_1 + C_2} \left[  \dd C_2 (e^{2 C_3} z_3  {-} z_1 z_2)+ \dd C_1 (e^{2 C_3} z_3 {+} z_1 z_2)\right] \hat{e}^{12} \nn
&+& \frac{2}{\Delta} e^{C_2 + C_3} \left[  \dd C_3 (e^{2 C_1} z_1 {-} z_2 z_3)+ \dd C_2 (e^{2 C_1} z_1 {+} z_2 z_3) \right] \hat{e}^{23} \nn
&-& \frac{1}{\Delta} \biggl[ A^1 e^{C_1} (z_2 e^{C_2}{+}z_1 z_3 e^{-C_2}) (e^{2 C_2}-e^{2C_3})+ A^2 e^{C_2} (e^{C_1}z_1{-}z_2 z_3 e^{-C_1})(e^{2 C_3}-e^{2 C_1})  \nn &{-}& A^3 e^{-C_1 -C_2} (e^{2C_2} z_2^2 - e^{2 C_1} z_1^2) (e^{2 C_1} - e^{2 C_2})\biggr] \hat{e}^{12} \nn
&-& \frac{1}{\Delta} \biggl[ A^2 e^{C_2} (z_3 e^{C_3} {+} z_1 z_2 e^{-C_3}) (e^{2 C_3}-e^{2C_1})+ A^3 e^{C_3} (e^{C_2}z_2 {-}z_1 z_3 e^{-C_2})(e^{2 C_1}-e^{2 C_2})  \nn &{-}& A^1 e^{-C_2 -C_3} (e^{2C_3} z_3^2 - e^{2 C_2} z_2^2) (e^{2 C_2} - e^{2 C_3})\biggr] \hat{e}^{23} \nn
&-& \frac{1}{\Delta} \biggl[ A^3 e^{C_3} (z_1 e^{C_1} {+} z_2 z_3 e^{-C_1}) (e^{2 C_1}-e^{2C_2})+ A^1 e^{C_1} (e^{C_3}z_3 {-}z_1 z_2 e^{-C_3})(e^{2 C_2}-e^{2 C_3})  \nn &{-}& A^2 e^{-C_3 -C_1} (e^{2C_1} z_1^2 - e^{2 C_3} z_3^2) (e^{2 C_3} - e^{2 C_1})\biggr] \hat{e}^{31} \nn
&+& \frac{1}{\Delta} \biggl[ {+} (e^{2 C_2+2 C_3}+ z_1^2) e^{C_1} \hat{e}^1 + (e^{2 C_3} z_3 {+}z_1 z_2)e^{C_2} \hat{e}^2 -(e^{2 C_2} z_2 {-} z_1 z_3) e^{C_3} \hat{e}^3 \biggr] e^{2 C_1} F^1 \nn
&+& \frac{1}{\Delta} \biggl[{+} (e^{2 C_3+2 C_1}+ z_2^2) e^{C_2} \hat{e}^2 + (e^{2 C_1} z_1 {+}z_2 z_3)e^{C_3} \hat{e}^3 -(e^{2 C_3} z_3 {-}z_1 z_2) e^{C_1} \hat{e}^1 \biggr] e^{2 C_2} F^2 \nn
&+& \frac{1}{\Delta} \biggl[ {+} (e^{2 C_1+2 C_2}+ z_3^2) e^{C_3} \hat{e}^3 + (e^{2 C_2} z_2 {+}z_1 z_3)e^{C_1} \hat{e}^1 -(e^{2 C_1} z_1 {-} z_2 z_3) e^{C_2} \hat{e}^2 \biggr] e^{2 C_3} F^3 \nonumber
\eea
\bea
&+& H \nn
&+& \frac{1}{\Delta} e^{2 C_1} B^1 \biggl[ (e^{2 C_2 + 2 C_3} -z_1^2) e^{-C_2 -C_3} \hat{e}^{23} + (-z_1 z_2 + z_3 e^{2 C_3} ) e^{-C_1 - C_3} \hat{e}^{31}\nn &-& (z_2 e^{2 C_2} +z_1 z_3) e^{-C_1 -C_2} \hat{e}^{12} \biggr] \nn
&+& \frac{1}{\Delta} e^{2 C_2} B^2 \biggl[ (e^{2 C_3 + 2 C_1} -z_2^2) e^{-C_1 -C_3} \hat{e}^{31} + (-z_2 z_3 + z_1 e^{2 C_1} ) e^{-C_1 - C_2} \hat{e}^{12}\nn &-& (z_3 e^{2 C_3} +z_1 z_2) e^{-C_2 -C_3} \hat{e}^{23} \biggr] \nn
&+& \frac{1}{\Delta} e^{2 C_3} B^3 \biggl[ (e^{2 C_1 + 2 C_2} -z_3^2) e^{-C_1 -C_2} \hat{e}^{12} + (-z_1 z_3 + z_2 e^{2 C_2} ) e^{-C_2 - C_3} \hat{e}^{23}\nn &-& (z_1 e^{2 C_1} +z_2 z_3) e^{-C_2 -C_3} \hat{e}^{31} \biggr] \nn
&-& \frac{1}{\Delta} \biggl[ - (z_2^2 e^{2 C_2} + z_3^2 e^{2 C_3}) e^{-C_1} \hat{e}^1+ (z_1 z_2 + z_3 e^{2 C_3}) e^{C_2} \hat{e}^2 + (z_1 z_3 - z_2 e^{2 C_2} ) e^{C_3} \hat{e}^3 \biggr] \DD B^1 \nn
&-&  \frac{1}{\Delta} \biggl[ - (z_1^2 e^{2 C_1} + z_3^2 e^{2 C_3}) e^{-C_2} \hat{e}^2+ (z_2 z_3 + z_1 e^{2 C_1}) e^{C_3} \hat{e}^3 + (z_1 z_2 - z_3 e^{2 C_3} ) e^{C_1} \hat{e}^1 \biggr] \DD B^2, \nn
&-& \frac{1}{\Delta} \biggl[ - (z_2^2 e^{2 C_2} + z_1^2 e^{2 C_1}) e^{-C_3} \hat{e}^3+ (z_1 z_3 + z_2 e^{2 C_2}) e^{C_1} \hat{e}^1 + (z_2 z_3 - z_1 e^{2 C_1} ) e^{C_2} \hat{e}^2 \biggr] \DD B^3. \nonumber
\eea
Here we have omitted wedge products for conciseness and the explicit form of $H$ can be found in (\ref{H0}). 

In our chosen frame, it is also helpful to record the derivative of the vielbein:
\bea
\label{spin1}
\dd \hat{e}^1 &=& - \dd C_1 \hat{e}^1 + \frac{2}{\Delta} \biggl[  (e^{2 C_2} z_2^2 + e^{2 C_3} z_3^2 ) {\dd C_1} \hat{e}^1+ e^{C_1 + C_2} (e^{2 C_3} z_3 - z_1 z_2) \dd C_2 \hat{e}^2 \nn &-& e^{C_1 +C_3} (e^{2 C_2} z_2 + z_1 z_3) \dd C_3 \hat{e}^3 \biggr] \nn
&+& \frac{1}{\Delta} \biggl[ A^2 (e^{2 C_1} - e^{2 C_3} )( e^{2 C_2} z_2 + z_1 z_3) + A^3 (e^{2 C_1} - e^{2 C_2} ) ( e^{2 C_3} z_3 - z_1 z_2)  \biggr] \hat{e}^1\nn
&+& \frac{1}{\Delta} \biggl[ -A^1 e^{C_1} (e^{2 C_2} - e^{2 C_3}) (e^{C_2} z_2 +e^{-C_2} z_1 z_3) - A^3 e^{C_1 - C_2} (z_1^2 + e^{2 C_2 + 2 C_3} ) (e^{2 C_1} - e^{2 C_2} )\biggr] \hat{e}^2 \nn
&+& \frac{1}{\Delta} \biggl[ A^1 e^{C_1} (e^{2 C_2} - e^{2 C_3}) (e^{C_3} z_3 -e^{-C_3} z_1 z_2) + A^2 e^{C_1- C_3} ( e^{2 C_2 + 2 C_3}+z_1^2) (e^{2 C_1} - e^{2 C_3} )\biggr] \hat{e}^3 \nn
&-& e^{C_1 - C_3} A^2 e^3 + e^{C_1 - C_2} A^3 \hat{e}^2 + \frac{e^{C_2}}{\Delta} (e^{C_2} z_2 + z_1 z_3) [1 + (e^{2 C_1} - e^{2 C_3}) e^{-2 C_2}] \hat{e}^{12} \nn &-& \frac{e^{C_3}}{\Delta} ( e^{2 C_3} z_3 - z_1 z_2) [1 + (e^{2 C_1} - e^{2 C_2}) e^{-2 C_3} ] \hat{e}^{31}\nn
&+& e^{C_1 -C_2 -C_3} [1 + \frac{1}{\Delta} ( e^{2 C_2} + e^{2 C_3} - e^{2 C_1}) (z_1^2 + e^{2 C_2 + 2 C_3}) ] \hat{e}^{23}\nonumber
\eea
\bea
&+& \frac{e^{C_1}}{\Delta} \biggl[ (e^{2 C_3} z_3^2 + e^{2 C_2} z_2^2) F^1 + e^{2 C_2} ( e^{2 C_3} z_3 - z_1 z_2) F^2 - e^{2 C_3} (e^{2 C_2} z_2 + z_1 z_3) F^3 \biggr] \nn
&+& \frac{e^{C_1}}{\Delta} \left[ (e^{2 C_2 + 2 C_3} +z_1^2) \DD B^1 + (-z_3 e^{2 C_3} + z_1 z_2 ) \DD B^2 + (z_2 e^{2 C_2} + z_1 z_3) \DD B^3 \right] \nn
&+& \frac{e^{C_1}}{\Delta} \biggl[ B^1 \{- (z_2 e ^{2 C_2} + z_1 z_3 ) e^{-C_2} \hat{e}^2 + (-z_3 e^{2 C_3} + z_1 z_2) e^{-C_3} \hat{e}^3 \}  \nn
&+& B^2 \{ (z_2 e ^{2 C_2} + z_1 z_3 ) e^{-C_1} \hat{e}^1 - (z_1^2 + e^{2 C_2 + 2 C_3} ) e^{-C_3} \hat{e}^3 \} \nn
&+& B^3 \{ (z_3 e ^{2 C_3} - z_1 z_2 ) e^{-C_1} \hat{e}^1 + (z_1^2  + e^{2 C_2 + 2 C_3}) e^{-C_2} \hat{e}^2 \}
\biggr].  
\eea
From here the spin-connection, $\omega$, can be worked out. First one reads of $c_{abc}$ from $ \dd e^a = - c_{abc} e^{b} \wedge e^c$, with the components of $\omega$, following from
\be
\omega_{abc} = \frac{1}{2} (c_{abc} + c_{bac} - c_{cab}). 
\ee
In determining $c_{abc}$ one should use the fact that expressions are cyclic under $1 \rightarrow 2 \rightarrow 3 \rightarrow 1$. 

We record some additional handy identities: 
\bea
\dd (\DD z_1) &=& \DD B^1 - \DD z_2 A^3 - z_2 F^3 + \DD z_3 A^2 +z_3 F^2, \nn
\sum_{a} z_a \DD z_a &=& \sum_{a} z_a e^{C_a} \hat{e}^{a}, \nn
\DD z_1 \DD z_2  \DD z_3 &=& \Delta e^{-C_1 -C_2 -C_3} \hat{e}^{123}, \nn
\dd (z_1 \DD z_2 \DD z_3) &=& \DD z_1 \DD z_2  \DD z_3 + (z_2 A^3-z_3 A^2) \DD z_2 \DD z_3 - z_1 \DD z_1 (\DD z_2 A^2 + \DD z_3 A^3) \nn
&+& z_1^2 \left( \DD z_3 F^3 + \DD z_2 F^2 \right) - z_1 \left( z_2 \DD z_2 +z_3 \DD z_3\right) F^1 \nn
&-& B^1 \DD z_2 \DD z_3 +z_1 ( \DD B^2 \DD z_3- \DD B^3 \DD z_2 ), \nn
\frac{1}{2} \epsilon_{abc} e^{2 C_a} z_a \DD z_b \DD z_c &=& \Delta \frac{1}{2} \epsilon_{abc} z_a e^{-C_b-C_c} \hat{e}^{bc}, 
\eea
where other expressions follow by exploiting cyclic symmetry in $1, 2, 3$. 

\section{Bianchi}
\label{sec:Bianchi} 

\setcounter{equation}{0}
In the body of the paper we have used the gravitino variation along the external seven-dimensional spacetime, $\delta \Psi_{\mu}$, to match the RR fields before and after T-duality. We noted that the T-dual RR flux (\ref{IIBfluxes}) is then consistent with the remaining supersymmetry conditions. As a further cross-check, it is prudent to check that the Bianchi identities before T-duality are mapped to the Bianchi identities post T-duality. 

With the original IIA RR flux Ansatz (\ref{IIAfluxes}), the Bianchi identities turn out to be
\bea
\label{Bianchi1} J^{a}_1 &=&m B^{a} + \DD K^{a}_0, \\
\label{Bianchi2} \dd G_2 &=& m H + J_1^a \wedge F^a, \\
\label{Bianchi3} \dd N_1 &=& K_0^a \DD B^a+ J_1^a \wedge B^a = \dd (K_0^a \, B^a), \\
\label{Bianchi4} \dd G_4 &=& H \wedge G_2 + L_3^{a} \wedge F^a, \\
\label{Bianchi5} L_{3}^{a} &=& \DD M_2^a - N_1 \wedge F^a + B^{a} \wedge G_2 - K_0^{a} H + \epsilon^{a}_{~bc} \DD B^b \wedge J_1^c, 
\eea
where repeated indices are summed. 

The upper two equations follow from $\dd F_2 = m H_3$, while the lower three equations are the result of $\dd F_4 = H_3 \wedge F_2$. There are two additional equations, but they are implied: 
\bea
\DD J^{a}_1 &=& m \DD B^{a} - \epsilon^{a}_{~bc} K_0^b F^c, \\
\DD L^{a}_3 &=& \epsilon^{a}_{~bc} F^b \wedge M^c_2 + H \wedge J^a_1 + \DD B^a \wedge G_2. 
\eea

In addition, it is useful to record the following equation of motion: 
\bea
\dd \left( e^{\sum_{a} C_a} *_7 G_4 \right) &=& - H \wedge N_1 - \DD B^a \wedge M_2^a + B^a \wedge L_3^a. 
\eea

We have checked that the Bianchi identities after T-duality are satisfied. The simplest example by far involves $\dd \hat{F}_1 =0$, which can be shown to hold by combining (\ref{Bianchi1}) and (\ref{Bianchi3}). This provides an extra check on the SU(2) transformation of the RR sector presented in section \ref{sec:IIAIIB}.

\end{document}